# Self-field, radiated energy, and radiated linear momentum of an accelerated point charge (Part 2)

Masud Mansuripur

*College of Optical Sciences, The University of Arizona, Tucson, Arizona, USA*

**Abstract**. Working within the framework of the classical theory of electrodynamics, we derive an exact mathematical solution to the problem of self-force (or radiation reaction) of an accelerated point-charge traveling in free space. In addition to deriving relativistic expressions for self electromagnetic fields, we obtain exact formulas for the rates of radiated energy and linear momentum *without* the need to renormalize the particle's mass – or to discard undesirable infinities. The relativistic expression of self-force known as the Abraham-Lorentz-Dirac equation is derived in two different ways. Certain properties of the self-force are examined, and an approximate formula for the self-force, first proposed by Landau and Lifshitz, is discussed in some detail.

**1. Introduction**. In Part 1 of the present paper [1], we showed that the scalar potential $\psi(\boldsymbol{r},t)$ of a point-charge $q$ traveling along the arbitrary trajectory $\boldsymbol{r}_p(t)$ is given by an integral containing a retarded as well as an advanced delta-function, namely,

Ignore this $\delta$-function at first, because its argument vanishes only when $t' > t$.

$$\psi(\boldsymbol{r},t) = \frac{q}{4\pi\varepsilon_0}\int_{t'=-\infty}^{t} |\boldsymbol{r}-\boldsymbol{r}_p(t')|^{-1}\{\delta[t-t'-|\boldsymbol{r}-\boldsymbol{r}_p(t')|/c] - \delta[t-t'+|\boldsymbol{r}-\boldsymbol{r}_p(t')|/c]\}\mathrm{d}t'. \quad (1)$$

The contribution of the advanced $\delta$-function is generally ignored, as its argument vanishes only when $t' > t$. We advocated, however, that when calculating the self-field acting on the point-particle, both the retarded and the advanced $\delta$-functions must be taken into account, each at one-half of its strength. This, in effect, is the mathematical underpinning of the idea proposed by Dirac in his seminal 1938 paper [2]. The retarded $\delta$-function's argument in Eq.(1) vanishes when $t' = t - |\boldsymbol{r}-\boldsymbol{r}_p(t')|/c$, at which point we will have

$$\partial\{t-t'-\sqrt{[\boldsymbol{r}-\boldsymbol{r}_p(t')]\cdot[\boldsymbol{r}-\boldsymbol{r}_p(t')]}/c\}/\partial t' = -1 + \dot{\boldsymbol{r}}_p(t')\cdot[\boldsymbol{r}-\boldsymbol{r}_p(t')]/c|\boldsymbol{r}-\boldsymbol{r}_p(t')|. \quad (2)$$

We thus arrive at the following closed form for the retarded scalar potential:

$$\psi(\boldsymbol{r},t) = \frac{q}{4\pi\varepsilon_0\{|\boldsymbol{r}-\boldsymbol{r}_p(t')| - [\boldsymbol{r}-\boldsymbol{r}_p(t')]\cdot\dot{\boldsymbol{r}}_p(t')/c\}}. \quad (3)$$

The corresponding expression for the retarded vector potential is similarly found to be

$$\boldsymbol{A}(\boldsymbol{r},t) = \frac{\mu_0 q\,\dot{\boldsymbol{r}}_p(t')}{4\pi\{|\boldsymbol{r}-\boldsymbol{r}_p(t')| - [\boldsymbol{r}-\boldsymbol{r}_p(t')]\cdot\dot{\boldsymbol{r}}_p(t')/c\}}. \quad (4)$$

Equations (3) and (4) are the standard Liénard-Wiechert potentials of a point-charge $q$ traveling along the arbitrary trajectory $\boldsymbol{r}_p(t)$ [3,4]. Denoting by $\boldsymbol{R} = \boldsymbol{r} - \boldsymbol{r}_p(t')$ the distance between the observation point $\boldsymbol{r}$ and the particle position $\boldsymbol{r}_p(t')$ at the retarded time $t'$, by $\hat{\boldsymbol{R}} = \boldsymbol{R}/R$ the unit-vector along the direction of $\boldsymbol{R}$, and by $\boldsymbol{\beta}(t') = \dot{\boldsymbol{r}}_p(t')/c$ and $\boldsymbol{\alpha}(t') = \ddot{\boldsymbol{r}}_p(t')/c^2$ the normalized velocity and acceleration of the particle at the retarded time $t'$, we find the electric and magnetic fields at the observation point $\boldsymbol{r}$ at time $t$ to be

$$\boldsymbol{E}(\boldsymbol{r},t) = -\boldsymbol{\nabla}\psi(\boldsymbol{r},t) - \partial_t\boldsymbol{A}(\boldsymbol{r},t) = \frac{q}{4\pi\varepsilon_0}\left[\frac{(1-\beta^2+\boldsymbol{R}\cdot\boldsymbol{\alpha})(\hat{\boldsymbol{R}}-\boldsymbol{\beta})}{R^2(1-\beta\cos\theta)^3} - \frac{\boldsymbol{\alpha}}{R(1-\beta\cos\theta)^2}\right]. \quad (5)$$

$$\boldsymbol{B}(\boldsymbol{r},t) = \boldsymbol{\nabla}\times\boldsymbol{A}(\boldsymbol{r},t) = \hat{\boldsymbol{R}}\times\boldsymbol{E}(\boldsymbol{r},t)/c. \quad (6)$$

Having determined the $\boldsymbol{E}$ and $\boldsymbol{B}$ fields at a distance $\boldsymbol{R}$ from the retarded position of the particle, it is rather straightforward (albeit tedious) to compute the radiation rates of electromagnetic (EM) energy and linear momentum, namely,

$$\dot{\mathcal{E}}(t) = \frac{2}{3}\left(\frac{q^2}{4\pi\varepsilon_0 c^3}\right)\left\{\frac{\ddot{r}_p^2(t)}{(1-\beta^2)^2} + \frac{[\dot{\boldsymbol{r}}_p(t)\cdot\ddot{\boldsymbol{r}}_p(t)]^2}{c^2(1-\beta^2)^3}\right\}. \quad (7)$$

$$\dot{\boldsymbol{p}}(t) = \frac{2}{3}\left(\frac{q^2}{4\pi\varepsilon_0 c^5}\right)\left\{\frac{\ddot{r}_p^2(t)}{(1-\beta^2)^2} + \frac{[\dot{\boldsymbol{r}}_p(t)\cdot\ddot{\boldsymbol{r}}_p(t)]^2}{c^2(1-\beta^2)^3}\right\}\dot{\boldsymbol{r}}_p(t). \tag{8}$$

Once again, Eqs.(7) and (8) are well-known results of classical electrodynamics, which have been derived and analyzed in standard textbooks [3-5]. In Ref.[1], we proceeded to compute the self $E$-field acting on the point-particle by integrating the $E$-field of Eq.(5) over a spherical surface of radius $R$ centered at the retarded position of the particle, allowing for contributions by both the advanced and the retarded $\delta$-functions that appear in Eq.(1). We then let $R \to 0$ and, in this limit, we found that the self $E$-field is given by

$$\begin{aligned}\boldsymbol{E}_{\text{self}}(t) = \frac{q}{8\pi\varepsilon_0 c^3}\Big\{&\left(\frac{1}{1-\beta^2} + \frac{1}{3} + \frac{1}{5}\beta^2 + \frac{1}{7}\beta^4 + \frac{1}{9}\beta^6 + \cdots\right)\dddot{\boldsymbol{r}}_p(t) \qquad \leftarrow \boxed{\boldsymbol{\beta}(t) = \dot{\boldsymbol{r}}_p(t)/c}\\ &+\frac{1}{c^2}\left[\frac{1}{1-\beta^2} - 3\left(\frac{1}{5} + \frac{1}{7}\beta^2 + \frac{1}{9}\beta^4 + \cdots\right)\right][\dot{\boldsymbol{r}}_p(t)\cdot\dddot{\boldsymbol{r}}_p(t) + \ddot{\boldsymbol{r}}_p(t)\cdot\ddot{\boldsymbol{r}}_p(t)]\dot{\boldsymbol{r}}_p(t)\\ &+\frac{1}{c^2}\left[\frac{4-2\beta^2}{(1-\beta^2)^2} - 6\left(\frac{1}{5} + \frac{1}{7}\beta^2 + \frac{1}{9}\beta^4 + \cdots\right)\right][\dot{\boldsymbol{r}}_p(t)\cdot\ddot{\boldsymbol{r}}_p(t)]\ddot{\boldsymbol{r}}_p(t)\\ &-\frac{1}{c^4}\left[\frac{1-3\beta^2}{(1-\beta^2)^2} - 15\left(\frac{1}{7} + \frac{1}{9}\beta^2 + \frac{1}{11}\beta^4 + \cdots\right)\right][\dot{\boldsymbol{r}}_p(t)\cdot\ddot{\boldsymbol{r}}_p(t)]^2\dot{\boldsymbol{r}}_p(t)\Big\}.\end{aligned} \tag{9}$$

Equation (9) was expected to be the *exact* relativistic expression of the self-field $\boldsymbol{E}$ acting on a point-charge $q$ moving in free space with instantaneous velocity $\dot{\boldsymbol{r}}_p(t)$, acceleration $\ddot{\boldsymbol{r}}_p(t)$, and time-rate-of-change of acceleration $\dddot{\boldsymbol{r}}_p(t)$ —the latter entity is often referred to as the "jerk" of the particle. At non-relativistic velocities, where $\beta$ is negligible, the self $E$-field of Eq.(9) reduces to

$$\boldsymbol{E}_{\text{self}}(t) \cong (4\pi\varepsilon_0)^{-1}(2q/3c^3)\dddot{\boldsymbol{r}}_p(t). \tag{10}$$

To our surprise, we found the self $E$-field of Eq.(9) to be *incorrect* at relativistic speeds of the particle, failing to yield the correct rate of EM energy radiation under certain circumstances [1]. We speculated that the root cause of the problem was the averaging of the self $E$-field over a (vanishingly small) *spherical* surface surrounding the particle, arguing that, if the particle at rest is modelled as a uniformly-charged solid sphere, then, when in motion, it would be contracted — perhaps assuming a spheroidal shape — in which case it would be inaccurate to average the $E$-field over a spherical surface centered at the particle's instantaneous location. In spite of this problem, which is manifested at relativistic speeds, the nonrelativistic self $E$-field of Eq.(10) remains valid. In particular, in the particle's instantaneous rest-frame, Eq.(10) becomes an exact expression of the self $E$-field. In the present paper, we arrive at the correct self-force acting on an accelerated point-charge via two different routes, and show the final results to be in complete agreement with each other, as well as with the well-known Abraham-Lorentz-Dirac equation [2,6,7]. Certain interesting properties of the self-force will be examined in Sections 4 through 8, and an approximate formula for self-force, first proposed by Landau and Lifshitz [5], will be discussed in some detail in Section 9.

**2. Deriving the self $\boldsymbol{E}$ and $\boldsymbol{B}$ fields from the rates of EM energy and linear momentum radiation**. As it turns out, the correct relativistic self-fields (both the $E$-field and the $B$-field) can be extracted directly from Eqs.(7) and (8), simply by rewriting these equations in a form that expresses the rate of radiation of energy as $q\boldsymbol{E}_{\text{self}}\cdot\dot{\boldsymbol{r}}_p(t)$, and the rate of radiation of linear momentum as $q[\boldsymbol{E}_{\text{self}} + \dot{\boldsymbol{r}}_p(t)\times\boldsymbol{B}_{\text{self}}]$.

We begin by noting that a *neutral* point-particle of mass $m$, following an arbitrary trajectory $\boldsymbol{r}_p(t)$ in free space, requires a driving force $\boldsymbol{f}(t) = \mathrm{d}\boldsymbol{p}(t)/\mathrm{d}t$, in accordance with the relativistic version of Newton's second law of motion. Here, $\boldsymbol{p}(t) = m\gamma(t)\boldsymbol{v}(t)$, where $\boldsymbol{v}(t) = \dot{\boldsymbol{r}}_p(t) = \mathrm{d}\boldsymbol{r}_p(t)/\mathrm{d}t$ is the particle's instantaneous velocity, and $\gamma(t) = [1-\beta^2(t)]^{-½}$ is the Lorentz-FitzGerald contraction/dilation factor written in terms of the normalized velocity $\boldsymbol{\beta}(t) = \boldsymbol{v}(t)/c$, with $c$ being the speed of light in vacuum. The mass-energy of the particle $\mathcal{E}(t) = \gamma(t)mc^2$, satisfies the relation $\dot{\mathcal{E}}(t) = \boldsymbol{f}(t)\cdot\boldsymbol{v}(t)$, where

$$\dot{\mathcal{E}}(t) = \dot{\gamma}(t)mc^2 = \gamma^3(t)\boldsymbol{\beta}(t)\cdot\dot{\boldsymbol{\beta}}(t)mc^2 = m\gamma^3(t)\boldsymbol{v}(t)\cdot\dot{\boldsymbol{v}}(t). \tag{11}$$

$$\begin{aligned}\boldsymbol{f}(t)\cdot\boldsymbol{v}(t) &= \dot{\boldsymbol{p}}(t)\cdot\boldsymbol{v}(t) = [\dot{\gamma}(t)m\boldsymbol{v}(t)+\gamma(t)m\dot{\boldsymbol{v}}(t)]\cdot\boldsymbol{v}(t)\\ &= m\big[\gamma^3(t)\boldsymbol{\beta}(t)\cdot\dot{\boldsymbol{\beta}}(t)\boldsymbol{v}(t)+\gamma(t)\dot{\boldsymbol{v}}(t)\big]\cdot\boldsymbol{v}(t)\\ &= m\gamma^3(t)\{[v(t)/c]^2+\gamma^{-2}(t)\}\boldsymbol{v}(t)\cdot\dot{\boldsymbol{v}}(t)\\ &= m\gamma^3(t)[\beta^2(t)+1-\beta^2(t)]\boldsymbol{v}(t)\cdot\dot{\boldsymbol{v}}(t)\\ &= m\gamma^3(t)\boldsymbol{v}(t)\cdot\dot{\boldsymbol{v}}(t). \end{aligned} \tag{12}$$

The time-rate-of-change of the angular momentum $\boldsymbol{\mathcal{L}}(t)=\boldsymbol{r}_p(t)\times\boldsymbol{p}(t)$ of the particle (relative to the origin of coordinates) is readily seen to be $\dot{\boldsymbol{\mathcal{L}}}(t)=\boldsymbol{v}(t)\times\boldsymbol{p}(t)+\boldsymbol{r}_p(t)\times\boldsymbol{f}(t)$, which equals the exerted torque $\boldsymbol{\mathcal{T}}(t)=\boldsymbol{r}_p(t)\times\boldsymbol{f}(t)$, since $\boldsymbol{v}(t)\times\boldsymbol{p}(t)=\gamma m\boldsymbol{v}(t)\times\boldsymbol{v}(t)=0$.

Suppose now that the point-particle of mass $m$ carries a charge $q$ as well. The point-charge, when accelerated, radiates EM energy and linear momentum at the respective rates of $\dot{\mathcal{E}}_{\text{EM}}(t)$ and $\dot{\boldsymbol{p}}_{\text{EM}}(t)$, given by Eqs.(7) and (8), which may be rewritten as follows:

$$\dot{\mathcal{E}}_{\text{EM}}(t)=\frac{2}{3}\left(\frac{q^2}{4\pi\varepsilon_0 c}\right)\left\{\frac{\dot{\boldsymbol{\beta}}(t)\cdot\dot{\boldsymbol{\beta}}(t)}{(1-\beta^2)^2}+\frac{[\boldsymbol{\beta}(t)\cdot\dot{\boldsymbol{\beta}}(t)]^2}{(1-\beta^2)^3}\right\}. \tag{13}$$

$$\dot{\boldsymbol{p}}_{\text{EM}}(t)=\frac{2}{3}\left(\frac{q^2}{4\pi\varepsilon_0 c^2}\right)\left\{\frac{\dot{\boldsymbol{\beta}}(t)\cdot\dot{\boldsymbol{\beta}}(t)}{(1-\beta^2)^2}+\frac{[\boldsymbol{\beta}(t)\cdot\dot{\boldsymbol{\beta}}(t)]^2}{(1-\beta^2)^3}\right\}\boldsymbol{\beta}(t)$$

$\boxed{\boldsymbol{A}\times(\boldsymbol{B}\times\boldsymbol{C})=(\boldsymbol{A}\cdot\boldsymbol{C})\boldsymbol{B}-(\boldsymbol{A}\cdot\boldsymbol{B})\boldsymbol{C}}\rightarrow$

$$=\frac{2}{3}\left(\frac{q^2}{4\pi\varepsilon_0 c^2}\right)\left\{\frac{\dot{\boldsymbol{\beta}}(t)\times[\boldsymbol{\beta}(t)\times\dot{\boldsymbol{\beta}}(t)]+[\boldsymbol{\beta}(t)\cdot\dot{\boldsymbol{\beta}}(t)]\dot{\boldsymbol{\beta}}(t)}{(1-\beta^2)^2}+\frac{[\boldsymbol{\beta}(t)\cdot\dot{\boldsymbol{\beta}}(t)]^2\boldsymbol{\beta}(t)}{(1-\beta^2)^3}\right\}. \tag{14}$$

We consider a situation in which the particle has an initial constant velocity $\boldsymbol{v}_1$ before $t_1$, and a final (possibly different) constant velocity $\boldsymbol{v}_2$ after $t_2>t_1$. In between $t_1$ and $t_2$, the particle follows an arbitrary trajectory $\boldsymbol{r}_p(t)$, which causes the radiation of EM energy and linear momentum at the rates of $\dot{\mathcal{E}}_{\text{EM}}(t)$ and $\dot{\boldsymbol{p}}_{\text{EM}}(t)$, respectively. The total emitted EM energy and momentum are thus given by $\int_{t_1}^{t_2}\dot{\mathcal{E}}_{\text{EM}}(t)\mathrm{d}t$ and $\int_{t_1}^{t_2}\dot{\boldsymbol{p}}_{\text{EM}}(t)\mathrm{d}t$. If we assume that the self-fields acting on the particle during the interval $(t_1,t_2)$ are $\boldsymbol{E}_{\text{self}}(t)$ and $\boldsymbol{B}_{\text{self}}(t)$, then it is reasonable to expect the total radiated energy to equal $-\int_{t_1}^{t_2}q\boldsymbol{E}_{\text{self}}(t)\cdot\boldsymbol{v}(t)\mathrm{d}t$, and also the total radiated momentum to equal $-\int_{t_1}^{t_2}q[\boldsymbol{E}_{\text{self}}(t)+\boldsymbol{v}(t)\times\boldsymbol{B}_{\text{self}}(t)]\mathrm{d}t$. Our goal in what follows is to extract expressions for $\boldsymbol{E}_{\text{self}}(t)$ and $\boldsymbol{B}_{\text{self}}(t)$ from Eqs.(13) and (14), consistent with these expectations. To this end, we write

$$\int_{t_1}^{t_2}\dot{\mathcal{E}}_{\text{EM}}(t)\mathrm{d}t=\frac{2}{3}\left(\frac{q^2}{4\pi\varepsilon_0 c}\right)\left\{\int_{t_1}^{t_2}\frac{\dot{\boldsymbol{\beta}}(t)\cdot\dot{\boldsymbol{\beta}}(t)}{(1-\beta^2)^2}\mathrm{d}t+\int_{t_1}^{t_2}\frac{[\boldsymbol{\beta}(t)\cdot\dot{\boldsymbol{\beta}}(t)]^2}{(1-\beta^2)^3}\mathrm{d}t\right\}$$

$$=\frac{2}{3}\left(\frac{q^2}{4\pi\varepsilon_0 c}\right)\left\{\underbrace{\frac{\dot{\boldsymbol{\beta}}(t)\cdot\boldsymbol{\beta}(t)}{(1-\beta^2)^2}\bigg|_{t_1}^{t_2}}_{0}-\int_{t_1}^{t_2}\underbrace{\left\{\frac{\ddot{\boldsymbol{\beta}}(t)}{(1-\beta^2)^2}+\frac{4[\boldsymbol{\beta}(t)\cdot\dot{\boldsymbol{\beta}}(t)]\dot{\boldsymbol{\beta}}(t)}{(1-\beta^2)^3}\right\}}_{\boldsymbol{E}_{\text{self}}^{(\text{i})}}\cdot\boldsymbol{\beta}(t)\mathrm{d}t+\int_{t_1}^{t_2}\frac{[\boldsymbol{\beta}(t)\cdot\dot{\boldsymbol{\beta}}(t)]^2}{(1-\beta^2)^3}\mathrm{d}t\right\}. \tag{15}$$

$\uparrow$ Integration by parts

In the above equation, the first term within the curly brackets vanishes because the velocities $\boldsymbol{v}(t_1)$ and $\boldsymbol{v}(t_2)$ are constant and, therefore, $\dot{\boldsymbol{\beta}}(t_1)=\dot{\boldsymbol{\beta}}(t_2)=0$. The integrand of the second term on the right-hand-side of Eq.(15) is in the form $\boldsymbol{E}(t)\cdot\boldsymbol{v}(t)$, which allows one to extract from the integrand a part of the self $E$-field identified as $\boldsymbol{E}_{\text{self}}^{(\text{i})}$. As for the third term, its integrand may likewise be written as $\boldsymbol{E}(t)\cdot\boldsymbol{v}(t)$, which would then yield the remaining part $\boldsymbol{E}_{\text{self}}^{(\text{ii})}$ of the self $E$-field. However, it is preferable to postpone the identification of $\boldsymbol{E}_{\text{self}}^{(\text{ii})}$ until after the next step.

Next, we use the expression for $\dot{\boldsymbol{p}}_{\text{EM}}(t)$ given in Eq.(14), and rearrange $\int_{t_1}^{t_2}\dot{\boldsymbol{p}}_{\text{EM}}(t)\mathrm{d}t$ in such a way as to enable the extraction of the self-force exerted on the particle, as follows:

$$\int_{t_1}^{t_2}\dot{\boldsymbol{p}}(t)\mathrm{d}t = \frac{2}{3}\left(\frac{q^2}{4\pi\varepsilon_0 c^2}\right)\left\{\int_{t_1}^{t_2}\frac{\dot{\boldsymbol{\beta}}(t)\times[\boldsymbol{\beta}(t)\times\dot{\boldsymbol{\beta}}(t)]}{(1-\beta^2)^2}\mathrm{d}t + \int_{t_1}^{t_2}\left\{\frac{[\boldsymbol{\beta}(t)\cdot\dot{\boldsymbol{\beta}}(t)]\dot{\boldsymbol{\beta}}(t)}{(1-\beta^2)^2}+\frac{[\boldsymbol{\beta}(t)\cdot\dot{\boldsymbol{\beta}}(t)]^2\boldsymbol{\beta}(t)}{(1-\beta^2)^3}\right\}\mathrm{d}t\right\}$$

$$= \frac{2}{3}\left(\frac{q^2}{4\pi\varepsilon_0 c^2}\right)\left\{\cancelto{0}{\frac{\boldsymbol{\beta}(t)\times[\boldsymbol{\beta}(t)\times\dot{\boldsymbol{\beta}}(t)]}{(1-\beta^2)^2}}\Bigg|_{t_1}^{t_2} - \int_{t_1}^{t_2}\boldsymbol{\beta}(t)\times\underbrace{\left\{\frac{\cancelto{0}{\dot{\boldsymbol{\beta}}(t)\times\dot{\boldsymbol{\beta}}(t)}+\boldsymbol{\beta}(t)\times\ddot{\boldsymbol{\beta}}(t)}{(1-\beta^2)^2}+\frac{4[\boldsymbol{\beta}(t)\cdot\dot{\boldsymbol{\beta}}(t)]\boldsymbol{\beta}(t)\times\dot{\boldsymbol{\beta}}(t)}{(1-\beta^2)^3}\right\}}_{\boldsymbol{B}_{\text{self}}}\mathrm{d}t\right.$$

Integration by parts

$$\left. + \int_{t_1}^{t_2}\underbrace{\left\{\frac{[\boldsymbol{\beta}(t)\cdot\dot{\boldsymbol{\beta}}(t)]\dot{\boldsymbol{\beta}}(t)}{(1-\beta^2)^2}+\frac{[\boldsymbol{\beta}(t)\cdot\dot{\boldsymbol{\beta}}(t)]^2\boldsymbol{\beta}(t)}{(1-\beta^2)^3}\right\}}_{-\boldsymbol{E}_{\text{self}}^{(\text{ii})}}\mathrm{d}t\right\}. \quad (16)$$

As before, the first term inside the curly brackets on the right-hand-side of Eq.(16) vanishes, since $\boldsymbol{v}(t_1)$ and $\boldsymbol{v}(t_2)$ are constant and, therefore, $\dot{\boldsymbol{\beta}}(t_1)=\dot{\boldsymbol{\beta}}(t_2)=0$. The integrand of the second term is in the form of $\boldsymbol{v}(t)\times\boldsymbol{B}(t)$, which yields $\boldsymbol{B}_{\text{self}}(t)$ as

$$\boldsymbol{B}_{\text{self}}(t) = \frac{2}{3}\left(\frac{q}{4\pi\varepsilon_0 c^3}\right)\left\{\frac{\boldsymbol{\beta}(t)\times\ddot{\boldsymbol{\beta}}(t)}{(1-\beta^2)^2}+\frac{4[\boldsymbol{\beta}(t)\cdot\dot{\boldsymbol{\beta}}(t)]\boldsymbol{\beta}(t)\times\dot{\boldsymbol{\beta}}(t)}{(1-\beta^2)^3}\right\}$$

$$= \frac{2}{3}\left(\frac{\mu_0 q}{4\pi c}\right)\boldsymbol{\beta}(t)\times\left\{\frac{\ddot{\boldsymbol{\beta}}(t)}{(1-\beta^2)^2}+\frac{4[\boldsymbol{\beta}(t)\cdot\dot{\boldsymbol{\beta}}(t)]\dot{\boldsymbol{\beta}}(t)}{(1-\beta^2)^3}\right\}. \quad (17)$$

Note that $\boldsymbol{B}_{\text{self}}(t)$ as given by Eq.(17) is orthogonal to the instantaneous velocity $\dot{\boldsymbol{r}}_p(t)=c\boldsymbol{\beta}(t)$ of the particle. This does *not* necessarily imply that the self $B$-field is perpendicular to the particle velocity; rather, it is a direct consequence of our method of extracting the self $B$-field that the projection of this field on $\dot{\boldsymbol{r}}_p(t)$ is automatically eliminated from the equations.

The integrand of the third term on the right-hand-side of Eq.(16) contains another part of the self $E$-field which we have marked as $-\boldsymbol{E}_{\text{self}}^{(\text{ii})}$, and which may be properly written as follows:

$$\boldsymbol{E}_{\text{self}}^{(\text{ii})} = -\frac{2}{3}\left(\frac{q}{4\pi\varepsilon_0 c^2}\right)\left\{\frac{[\boldsymbol{\beta}(t)\cdot\dot{\boldsymbol{\beta}}(t)]\dot{\boldsymbol{\beta}}(t)}{(1-\beta^2)^2}+\frac{[\boldsymbol{\beta}(t)\cdot\dot{\boldsymbol{\beta}}(t)]^2\boldsymbol{\beta}(t)}{(1-\beta^2)^3}\right\}. \quad (18)$$

The EM energy drawn from the moving point-particle by the action of $\boldsymbol{E}_{\text{self}}^{(\text{ii})}$ is readily seen to be

$$\int_{t_1}^{t_2} q\boldsymbol{E}_{\text{self}}^{(\text{ii})}(t)\cdot\boldsymbol{v}(t)\mathrm{d}t = -\frac{2}{3}\left(\frac{q^2}{4\pi\varepsilon_0 c}\right)\int_{t_1}^{t_2}\left\{\frac{[\boldsymbol{\beta}(t)\cdot\dot{\boldsymbol{\beta}}(t)]\dot{\boldsymbol{\beta}}(t)}{(1-\beta^2)^2}+\frac{[\boldsymbol{\beta}(t)\cdot\dot{\boldsymbol{\beta}}(t)]^2\boldsymbol{\beta}(t)}{(1-\beta^2)^3}\right\}\cdot\boldsymbol{\beta}(t)\mathrm{d}t$$

$$= -\frac{2}{3}\left(\frac{q^2}{4\pi\varepsilon_0 c}\right)\int_{t_1}^{t_2}\left\{\frac{[\boldsymbol{\beta}(t)\cdot\dot{\boldsymbol{\beta}}(t)]^2}{(1-\beta^2)^2}+\frac{[\boldsymbol{\beta}(t)\cdot\dot{\boldsymbol{\beta}}(t)]^2\beta^2(t)}{(1-\beta^2)^3}\right\}\mathrm{d}t = -\frac{2}{3}\left(\frac{q^2}{4\pi\varepsilon_0 c}\right)\int_{t_1}^{t_2}\frac{[\boldsymbol{\beta}(t)\cdot\dot{\boldsymbol{\beta}}(t)]^2}{(1-\beta^2)^3}\mathrm{d}t. \quad (19)$$

The above result is identical to the last term on the right-hand-side of Eq.(15). Thus, the total self $E$-field must be the sum of $\boldsymbol{E}_{\text{self}}^{(\text{i})}$ and $\boldsymbol{E}_{\text{self}}^{(\text{ii})}$. The only point that remains to be verified is that the addition of $\boldsymbol{E}_{\text{self}}^{(\text{i})}$ to $\boldsymbol{E}_{\text{self}}^{(\text{ii})}$ will not adversely affect Eq.(16). This is because the integral from $t_1$ to $t_2$ of $\boldsymbol{E}_{\text{self}}^{(\text{i})}$ equals zero, that is,

$$\int_{t_1}^{t_2}\boldsymbol{E}_{\text{self}}^{(\text{i})}\mathrm{d}t = \frac{2}{3}\left(\frac{q}{4\pi\varepsilon_0 c^2}\right)\int_{t_1}^{t_2}\left\{\frac{\ddot{\boldsymbol{\beta}}(t)}{(1-\beta^2)^2}+\frac{4[\boldsymbol{\beta}(t)\cdot\dot{\boldsymbol{\beta}}(t)]\dot{\boldsymbol{\beta}}(t)}{(1-\beta^2)^3}\right\}\mathrm{d}t = \frac{2}{3}\left(\frac{q}{4\pi\varepsilon_0 c^2}\right)\frac{\dot{\boldsymbol{\beta}}(t)}{(1-\beta^2)^2}\Bigg|_{t_1}^{t_2} = 0. \quad (20)$$

The complete expression of the self $E$-field may finally be written as follows:

$$\boldsymbol{E}_{\text{self}}(t) = \boldsymbol{E}_{\text{self}}^{(\text{i})}(t)+\boldsymbol{E}_{\text{self}}^{(\text{ii})}(t)$$

$$= \frac{2}{3}\left(\frac{q}{4\pi\varepsilon_0 c^2}\right)\left\{\frac{\ddot{\boldsymbol{\beta}}(t)}{(1-\beta^2)^2}+\frac{4[\boldsymbol{\beta}(t)\cdot\dot{\boldsymbol{\beta}}(t)]\dot{\boldsymbol{\beta}}(t)}{(1-\beta^2)^3}-\frac{[\boldsymbol{\beta}(t)\cdot\dot{\boldsymbol{\beta}}(t)]\dot{\boldsymbol{\beta}}(t)}{(1-\beta^2)^2}-\frac{[\boldsymbol{\beta}(t)\cdot\dot{\boldsymbol{\beta}}(t)]^2\boldsymbol{\beta}(t)}{(1-\beta^2)^3}\right\}. \quad (21)$$

Combining Eqs.(17) and (21), the total self-force acting on the point-particle is found to be

$$
\begin{aligned}
\boldsymbol{f}_{\text{self}}(t) &= q\boldsymbol{E}_{\text{self}}(t) + q\boldsymbol{v}(t)\times\boldsymbol{B}_{\text{self}}(t)\\
&= \frac{2}{3}\left(\frac{q^2}{4\pi\varepsilon_0 c^2}\right)\left[\frac{\ddot{\boldsymbol{\beta}}(t)}{(1-\beta^2)^2} + \frac{4[\boldsymbol{\beta}(t)\cdot\dot{\boldsymbol{\beta}}(t)]\dot{\boldsymbol{\beta}}(t)}{(1-\beta^2)^3} - \frac{[\boldsymbol{\beta}(t)\cdot\dot{\boldsymbol{\beta}}(t)]\dot{\boldsymbol{\beta}}(t)}{(1-\beta^2)^2} - \frac{[\boldsymbol{\beta}(t)\cdot\dot{\boldsymbol{\beta}}(t)]^2\boldsymbol{\beta}(t)}{(1-\beta^2)^3}\right]\\
&\quad + \frac{2}{3}\left(\frac{q^2}{4\pi\varepsilon_0 c^2}\right)\boldsymbol{\beta}(t)\times\left[\frac{\boldsymbol{\beta}(t)\times\ddot{\boldsymbol{\beta}}(t)}{(1-\beta^2)^2} + \frac{4[\boldsymbol{\beta}(t)\cdot\dot{\boldsymbol{\beta}}(t)]\boldsymbol{\beta}(t)\times\dot{\boldsymbol{\beta}}(t)}{(1-\beta^2)^3}\right]\\
&= \frac{2}{3}\left(\frac{q^2}{4\pi\varepsilon_0 c^2}\right)\left\{\frac{\ddot{\boldsymbol{\beta}}(t)}{(1-\beta^2)^2} + \frac{4[\boldsymbol{\beta}(t)\cdot\dot{\boldsymbol{\beta}}(t)]\dot{\boldsymbol{\beta}}(t)}{(1-\beta^2)^3} - \frac{[\boldsymbol{\beta}(t)\cdot\dot{\boldsymbol{\beta}}(t)]\dot{\boldsymbol{\beta}}(t)}{(1-\beta^2)^2} - \frac{[\boldsymbol{\beta}(t)\cdot\dot{\boldsymbol{\beta}}(t)]^2\boldsymbol{\beta}(t)}{(1-\beta^2)^3}\right.\\
&\qquad\left. + \left[\frac{\boldsymbol{\beta}(t)\cdot\ddot{\boldsymbol{\beta}}(t)}{(1-\beta^2)^2} + \frac{4[\boldsymbol{\beta}(t)\cdot\dot{\boldsymbol{\beta}}(t)]^2}{(1-\beta^2)^3}\right]\boldsymbol{\beta}(t) - \left[\frac{\ddot{\boldsymbol{\beta}}(t)}{(1-\beta^2)^2} + \frac{4[\boldsymbol{\beta}(t)\cdot\dot{\boldsymbol{\beta}}(t)]\dot{\boldsymbol{\beta}}(t)}{(1-\beta^2)^3}\right]\beta^2(t)\right\}\\
&= \frac{2}{3}\left(\frac{q^2}{4\pi\varepsilon_0 c^2}\right)\left\{\frac{\ddot{\boldsymbol{\beta}}(t)}{1-\beta^2} + \frac{[\boldsymbol{\beta}(t)\cdot\ddot{\boldsymbol{\beta}}(t)]\boldsymbol{\beta}(t)}{(1-\beta^2)^2} + \frac{3[\boldsymbol{\beta}(t)\cdot\dot{\boldsymbol{\beta}}(t)]\dot{\boldsymbol{\beta}}(t)}{(1-\beta^2)^2} + \frac{3[\boldsymbol{\beta}(t)\cdot\dot{\boldsymbol{\beta}}(t)]^2\boldsymbol{\beta}(t)}{(1-\beta^2)^3}\right\}\\
&= \frac{2}{3}\left(\frac{q^2}{4\pi\varepsilon_0 c^2}\right)\left\{\frac{\ddot{\boldsymbol{\beta}}(t)}{1-\beta^2} + \frac{\boldsymbol{\beta}(t)\times[\boldsymbol{\beta}(t)\times\ddot{\boldsymbol{\beta}}(t)]+\beta^2\ddot{\boldsymbol{\beta}}(t)}{(1-\beta^2)^2} + \frac{3[\boldsymbol{\beta}(t)\cdot\dot{\boldsymbol{\beta}}(t)]\{(1-\beta^2)\dot{\boldsymbol{\beta}}(t)+[\boldsymbol{\beta}(t)\cdot\dot{\boldsymbol{\beta}}(t)]\boldsymbol{\beta}(t)\}}{(1-\beta^2)^3}\right\}\\
&= \frac{2}{3}\left(\frac{q^2}{4\pi\varepsilon_0 c^2}\right)\left\{\frac{\ddot{\boldsymbol{\beta}}(t) + [\ddot{\boldsymbol{\beta}}(t)\times\boldsymbol{\beta}(t)]\times\boldsymbol{\beta}(t)}{(1-\beta^2)^2} + \frac{3[\boldsymbol{\beta}(t)\cdot\dot{\boldsymbol{\beta}}(t)]\{\dot{\boldsymbol{\beta}}(t) + [\dot{\boldsymbol{\beta}}(t)\times\boldsymbol{\beta}(t)]\times\boldsymbol{\beta}(t)\}}{(1-\beta^2)^3}\right\}.
\end{aligned}
\tag{22}
$$

The externally-applied force needed to maintain the point-particle of mass $m$ and charge $q$ on its trajectory $\boldsymbol{r}_p(t)$ must, therefore, be the sum of the force $\boldsymbol{f}_1(t) = \dot{\boldsymbol{p}}(t) = \partial\left[mc\boldsymbol{\beta}(t)/\sqrt{1-\beta^2(t)}\,\right]/\partial t$, which is needed to accelerate/decelerate a *neutral* particle of mass $m$ in accordance with the relativistic version of Newton's second law of motion, and a second force, $\boldsymbol{f}_2(t) = -\boldsymbol{f}_{\text{self}}(t)$, that balances the effects of $\boldsymbol{f}_{\text{self}}(t)$.

---

**Digression**: There exist other terms that one could add to the self-force expression in Eq.(22) without affecting either $\int_{t_1}^{t_2}\boldsymbol{f}_{\text{self}}(t)\mathrm{d}t$ or $\int_{t_1}^{t_2}\boldsymbol{f}_{\text{self}}(t)\cdot\boldsymbol{v}(t)\mathrm{d}t$. For instance, any force that is proportional to $\dddot{\boldsymbol{\beta}}(t)$ will have the property that $\int_{t_1}^{t_2}\dddot{\boldsymbol{\beta}}(t)\mathrm{d}t = \ddot{\boldsymbol{\beta}}(t_2) - \ddot{\boldsymbol{\beta}}(t_1) = 0$, and also

$$\int_{t_1}^{t_2}\dddot{\boldsymbol{\beta}}(t)\cdot\boldsymbol{v}(t)\mathrm{d}t = \cancelto{0}{c\ddot{\boldsymbol{\beta}}(t)\cdot\boldsymbol{\beta}(t)\Big|_{t_1}^{t_2}} - c\int_{t_1}^{t_2}\ddot{\boldsymbol{\beta}}(t)\cdot\dot{\boldsymbol{\beta}}(t)\mathrm{d}t = \cancelto{0}{-c\dot{\boldsymbol{\beta}}(t)\cdot\dot{\boldsymbol{\beta}}(t)\Big|_{t_1}^{t_2}} + c\int_{t_1}^{t_2}\dot{\boldsymbol{\beta}}(t)\cdot\ddot{\boldsymbol{\beta}}(t)\mathrm{d}t. \tag{23}$$

The expression on the right-hand side of Eq.(23) is seen to be equal to its own negative, which is possible only if the expression vanishes. It is also easy to show that a force term that is proportional to $\dddot{\boldsymbol{\beta}}(t)$ does not impart a net angular momentum to the particle during the time interval $(t_1, t_2)$. This is because the integrated torque, being proportional to $\int_{t_1}^{t_2}\boldsymbol{r}_p(t)\times\dddot{\boldsymbol{\beta}}(t)\mathrm{d}t$, vanishes as follows:

$$\int_{t_1}^{t_2}\boldsymbol{r}_p(t)\times\dddot{\boldsymbol{\beta}}(t)\mathrm{d}t = \cancelto{0}{\boldsymbol{r}_p(t)\times\ddot{\boldsymbol{\beta}}(t)\Big|_{t_1}^{t_2}} - c\int_{t_1}^{t_2}\boldsymbol{\beta}(t)\times\ddot{\boldsymbol{\beta}}(t)\mathrm{d}t = \cancelto{0}{-c\boldsymbol{\beta}(t)\times\dot{\boldsymbol{\beta}}(t)\Big|_{t_1}^{t_2}} + c\int_{t_1}^{t_2}\dot{\boldsymbol{\beta}}(t)\times\dot{\boldsymbol{\beta}}(t)\mathrm{d}t = 0. \tag{24}$$

Consequently, one cannot rule out the possibility that the self-force of Eq.(22) could have an additional term proportional to $\dddot{\boldsymbol{\beta}}(t)$. Similar arguments can be made for any force that is proportional to an odd derivative of $\boldsymbol{\beta}(t)$ of higher-order, such as the $5^{\text{th}}, 7^{\text{th}}, 9^{\text{th}}, \cdots$ derivative.

One may be tempted to consider other possibilities for the self-force as well. For instance, a force term that is proportional to $\boldsymbol{\beta}(t)\times\ddot{\boldsymbol{\beta}}(t)$ does not impart a net energy to the particle, since its dot-product with the particle velocity is always zero, that is, $[\boldsymbol{\beta}(t)\times\ddot{\boldsymbol{\beta}}(t)]\cdot\boldsymbol{\beta}(t) = 0$. Such a force cannot impart a net momentum to the particle either, simply because its integral is given by

$$\int_{t_1}^{t_2}\boldsymbol{\beta}(t)\times\ddot{\boldsymbol{\beta}}(t)\mathrm{d}t = \cancelto{0}{\boldsymbol{\beta}(t)\times\dot{\boldsymbol{\beta}}(t)\Big|_{t_1}^{t_2}} - \int_{t_1}^{t_2}\dot{\boldsymbol{\beta}}(t)\times\dot{\boldsymbol{\beta}}(t)\mathrm{d}t = 0. \tag{25}$$

However, the net angular momentum imparted to the point-particle by a force that is proportional to $\boldsymbol{\beta}(t)\times\ddot{\boldsymbol{\beta}}(t)$ does *not* automatically vanish, because the following integral is *not* necessarily zero:

$$\int_{t_1}^{t_2}\boldsymbol{r}_p(t)\times[\boldsymbol{\beta}(t)\times\ddot{\boldsymbol{\beta}}(t)]\mathrm{d}t = \cancelto{0}{\boldsymbol{r}_p(t)\times[\boldsymbol{\beta}(t)\times\dot{\boldsymbol{\beta}}(t)]_{t_1}^{t_2}} - c\int_{t_1}^{t_2}\boldsymbol{\beta}(t)\times[\boldsymbol{\beta}(t)\times\dot{\boldsymbol{\beta}}(t)]\mathrm{d}t. \tag{26}$$

Consequently, a term proportional to $\boldsymbol{\beta}(t)\times\ddot{\boldsymbol{\beta}}(t)$ *cannot* be added to the self-force expression appearing in Eq.(22).

---

**3. Deriving the self-force by Lorentz transformation from the particle's rest frame to the lab frame**. According to Eq.(10), the exact self-force of the EM field acting on the charged particle in its own rest-frame (r-f) is given by $q^2\ddot{\boldsymbol{\beta}}_{\text{r-f}}/(6\pi\varepsilon_0 c^2)$, where $\ddot{\boldsymbol{r}}_p(t) = c\dot{\boldsymbol{\beta}}_{\text{r-f}}$; this result is an immediate consequence of Eq.(9). In the special theory of relativity, one must follow certain simple procedures to construct a 4-vector out of the time-derivative of an existing 4-vector. Thus, for instance, the 4-velocity $\mathbb{v}(t)$ is obtained by differentiating the spacetime coordinates $[ct, \boldsymbol{r}_p(t)]$ of a point-particle. Similarly, the 4-acceleration $\mathbb{a}(t)$ is obtained by differentiating the 4-velocity $\mathbb{v}(t)$, and the 4-jerk is obtained by differentiating the 4-acceleration. Appendix A describes in some detail the rules and procedures for computing the time-derivatives of various 4-vectors. Following these rules, we determine $\ddot{\boldsymbol{\beta}}_{\text{r-f}}$ in terms of $\boldsymbol{\beta}$, $\dot{\boldsymbol{\beta}}$, and $\ddot{\boldsymbol{\beta}}$ in the laboratory frame. The corresponding 4-force may then be Lorentz transformed back into the laboratory frame. In the process, the component of the 4-force that is perpendicular to $\dot{\boldsymbol{r}}_p(t)$ remains intact, while its parallel component gets multiplied by $\gamma(t)$. Once in the laboratory frame, one can recover the 3-force by dividing both components of the 4-force (i.e., its $\parallel$ and $\perp$ components) by $\gamma(t)$. Subsequently, the self-force acting on the point-particle in the laboratory frame is found to be (see Appendix A, Eq.(A15))

$$\boldsymbol{f}_{\text{self}}(t) = \left(\frac{q^2}{6\pi\varepsilon_0 c^2}\right)\left(\ddot{\boldsymbol{\beta}}_{\text{r-f}}^{\parallel} + \gamma^{-1}\ddot{\boldsymbol{\beta}}_{\text{r-f}}^{\perp}\right) = \left(\frac{q^2}{6\pi\varepsilon_0 c^2}\right)\left[\gamma^2\ddot{\boldsymbol{\beta}} + \gamma^4(\boldsymbol{\beta}\cdot\ddot{\boldsymbol{\beta}})\boldsymbol{\beta} + 3\gamma^4(\boldsymbol{\beta}\cdot\dot{\boldsymbol{\beta}})\dot{\boldsymbol{\beta}} + 3\gamma^6(\boldsymbol{\beta}\cdot\dot{\boldsymbol{\beta}})^2\boldsymbol{\beta}\right]. \quad (27)$$

Equation (27), which is the exact relativistic formula for the self-force acting on an accelerated point-charge, is the well-known Abraham-Lorentz-Dirac equation [2,6,7], which is also in complete agreement with the result that we obtained indirectly in Eq.(22).

**4. Launching the particle on a pre-determined trajectory**. One may now argue that a point-particle of mass $m$ and charge $q$ can be launched on an arbitrary spacetime trajectory $\boldsymbol{r}_p(t)$, in the following way: As a first step, one ignores the charge $q$ of the particle, and uses its relativistic momentum $\boldsymbol{p}(t) = \gamma m\boldsymbol{v}(t) = \gamma m\dot{\boldsymbol{r}}_p(t)$ to compute the force $\boldsymbol{f}_1(t) = \dot{\boldsymbol{p}}(t)$ that needs to be exerted on the particle on account of its inertial mass $m$. Next, one computes the self-force from Eq.(27) and proceeds to apply an additional force $\boldsymbol{f}_2(t) = -\boldsymbol{f}_{\text{self}}(t)$ to the particle. This additional force supplies the radiated energy as well as the radiated momentum — which are lost to the particle in consequence of its accelerated motion — but $\boldsymbol{f}_2(t)$ does not alter the trajectory $\boldsymbol{r}_p(t)$ of the particle, which, in the absence of radiation, is controlled solely by $\boldsymbol{f}_1(t)$. There may be periods of time during which the particle radiates EM energy and/or momentum while the instantaneous (externally-applied) force $\boldsymbol{f}_2(t)$, and/or the dot-product $\boldsymbol{f}_2(t)\cdot\boldsymbol{v}(t)$, will be zero. Such a situation might arise, for instance, at non-relativistic velocities when the particle has a constant acceleration, that is, $\ddot{\boldsymbol{\beta}}(t) = 0$; see Eq.(27). During such times, the radiated energy/momentum will be drawn from the intrinsic EM field energy/momentum of the charged-particle, of which there is an infinite supply in the space surrounding the zero-size particle. Needless to say, the particle's intrinsic EM energy/momentum will be automatically restored by the time the particle attains a constant final velocity at a later time.

**5. Self-acceleration of a point-charge in vacuum**. In the absence of external forces, the equation of motion of a point-charge moving along the $x$-axis under the influence of its self-force $\boldsymbol{f}_{\text{self}}(t)$ alone, can be written in terms of the particle's proper time $s$. This is achieved by first expressing the particle's position $x(t)$, velocity $\dot{x}(t) = c\beta(t)$, acceleration $\ddot{x}(t) = c\dot{\beta}(t)$, and jerk $\dddot{x}(t) = c\ddot{\beta}(t)$, as functions of the proper time $s$; that is,

$$\dot{x}(s) = (\mathrm{d}t/\mathrm{d}s)\dot{x}(t) = \gamma(t)\dot{x}(t) \quad \rightarrow \quad \beta(s) = \gamma(t)\beta(t). \quad (28)$$

$$\beta^2(s) = \beta^2(t)/[1-\beta^2(t)] \;\rightarrow\; \beta^2(t) = \beta^2(s)/[1+\beta^2(s)] \;\rightarrow\; \gamma^2(t) = 1+\beta^2(s). \quad (29)$$

$$\dot{\beta}(s) = \gamma(t)\frac{\mathrm{d}}{\mathrm{d}t}[\gamma(t)\beta(t)] = \gamma(t)[\dot{\gamma}(t)\beta(t) + \gamma(t)\dot{\beta}(t)] = \gamma^4\beta^2\dot{\beta} + \gamma^2\dot{\beta} = \gamma^4(t)\dot{\beta}(t). \quad (30)$$

$$\ddot{\beta}(s) = \gamma(t)\frac{\mathrm{d}}{\mathrm{d}t}[\gamma^4(t)\dot{\beta}(t)] = 4\gamma^7(t)\beta(t)\dot{\beta}^2(t) + \gamma^5(t)\ddot{\beta}(t). \quad (31)$$

Expressed as a function of the proper time $s$, the self-force $f_{\text{self}}(t)\hat{\boldsymbol{x}}$ of Eq.(27) is now written as follows:

$$f_{\text{self}}(t) = \left(\frac{q^2}{6\pi\varepsilon_0 c^2}\right)\left(3\gamma^6\beta^3\dot{\beta}^2 + \gamma^4\beta^2\ddot{\beta} + 3\gamma^4\beta\dot{\beta}^2 + \gamma^2\ddot{\beta}\right)$$
$$= \left(\frac{q^2}{6\pi\varepsilon_0 c^2}\right)\left[3\gamma^6(t)\beta(t)\dot{\beta}^2(t) + \gamma^4(t)\ddot{\beta}(t)\right] = \left(\frac{q^2}{6\pi\varepsilon_0 c^2}\right)\left[\ddot{\beta}(s) - \frac{\beta(s)\dot{\beta}^2(s)}{\gamma^2(t)}\right]\gamma^{-1}(t)$$
$$= \left(\frac{q^2}{6\pi\varepsilon_0 c^2}\right)\left[\ddot{\beta}(s) - \frac{\beta(s)\dot{\beta}^2(s)}{1+\beta^2(s)}\right]\gamma^{-1}(t). \qquad (32)$$

Equating the driving force $f_{\text{self}}(t)\hat{\boldsymbol{x}}$ with the time-rate of change of the linear momentum of the particle, and invoking Eq.(30), we are now in a position to write the equation of motion of the particle along the $x$-axis, as follows:

$$f_{\text{self}}(t) = \frac{\mathrm{d}}{\mathrm{d}t}[m\gamma(t)\dot{x}(t)] = mc[\dot{\gamma}(t)\beta(t) + \gamma(t)\dot{\beta}(t)] = mc\gamma^3(t)\dot{\beta}(t) = mc\gamma^{-1}(t)\dot{\beta}(s). \qquad (33)$$

Finally, comparing Eqs.(32) and (33), and defining the constant parameter $a = 6\pi\varepsilon_0 mc^3/q^2$, we arrive at the desired equation of motion in terms of the proper time $s$, namely,

$$a\dot{\beta}(s) - \ddot{\beta}(s) + \frac{\beta(s)\dot{\beta}^2(s)}{1+\beta^2(s)} = 0. \qquad (34)$$

To solve this equation, we resort to the change of variable $\beta(s) = \sinh[\zeta(s)]$ and rewrite Eq.(34) as follows:

$$a\dot{\zeta}\cosh(\zeta) - \ddot{\zeta}\cosh(\zeta) - \dot{\zeta}^2\sinh(\zeta) + \frac{\dot{\zeta}^2\cosh^2(\zeta)\sinh(\zeta)}{1+\sinh^2(\zeta)} = 0 \quad \rightarrow \quad \ddot{\zeta} = a\dot{\zeta}. \qquad (35)$$

The solution of Eq.(35) is readily seen to be $\zeta(s) = \zeta_0 + \zeta_1 e^{as}$, where $\zeta_0$ and $\zeta_1$ are arbitrary integration constants. We thus find $\beta(s) = \sinh(\zeta_0 + \zeta_1 e^{as})$, with $\zeta_0$ and $\zeta_1$ to be determined by the initial velocity and acceleration of the particle at $s = -\infty$. (This is the same solution as given in Eq.(30) of Dirac's paper [2].)

**6. Runaway solutions**. The main trouble with the self-force of Eq.(27) is that it admits runaway solutions [2]. The simplest example of such solutions occurs at non-relativistic velocities, when one can approximate Eq.(27) as $\boldsymbol{f}_{\text{self}}(t) \cong q^2\ddot{\boldsymbol{v}}(t)/(6\pi\varepsilon_0 c^3)$. Assuming that the particle is initially at rest during $t \le 0$, and that, after $t = 0$, an external force $f_{\text{ext}}(t)\hat{\boldsymbol{x}}$ is applied along the $x$-axis, the non-relativistic equation of motion of the particle, namely, $m\dot{\boldsymbol{v}}(t) = \boldsymbol{f}_{\text{self}}(t) + \boldsymbol{f}_{\text{ext}}(t)$, can be simplified for this one-dimensional motion along the $x$-axis, as follows:

$$\dot{v}(t) - \tau_0\ddot{v}(t) = f_{\text{ext}}(t)/m. \qquad (36)$$

In this equation, $\tau_0 = q^2/(6\pi\varepsilon_0 mc^3)$ is a constant coefficient that, assuming the particle under consideration is an electron, equals ⅔ the time it takes light to travel across its classical diameter.[†] Note that any constant velocity $v_0$ that the particle may have during $t \le 0$ can be added to the solution $v(t)$ of Eq.(36), simply because $v(t) = v_0$ is a homogeneous solution of the differential equation. Another homogeneous solution is $v(t) = v_1\exp(t/\tau_0)$, which we will have occasion to utilize shortly.

The impulse-response of the linear time-invariant system governed by Eq.(36) is obtained by solving the differential equation when its right-hand-side is set equal to $\delta(t)$. The impulse-response is readily found to be

$$v_{\text{i.r.}}(t) = \text{step}(t)[1 - \exp(t/\tau_0)]. \qquad (37)$$

Note that, at $t = 0$, this impulse-response is continuous, but its derivative, $\dot{v}_{\text{i.r.}}(t)$, has a discontinuity (or jump) equal to $-1/\tau_0$. This jump, of course, is necessary if the term $-\tau_0\ddot{v}(t)$ in Eq.(36) is to reproduce the $\delta$-function appearing on the right-hand side of the equation. The full solution of Eq.(36) is thus obtained as a convolution of the excitation function $f_{\text{ext}}(t)/m$ and the impulse-response given by Eq.(37), as follows:

---

[†] Let a spherical shell of radius $r_0$ and charge $q$, where the charge is uniformly distributed over the sphere's surface, be a model for a stationary electron. Upon integration over the entire space, the $E$-field energy-density $\frac{1}{2}\varepsilon_0 E^2(r) = q^2/(32\pi^2\varepsilon_0 r^4)$ outside the shell yields the total EM energy of the electron as $\mathcal{E} = q^2/(8\pi\varepsilon_0 r_0)$. Equating $\mathcal{E}$ to the mass-energy $mc^2$, one obtains the *classical diameter* of the electron as $2r_0 = q^2/(4\pi\varepsilon_0 mc^2)$.

$$v(t) = v_{\text{i.r.}}(t) * f_{\text{ext}}(t)/m = (1/m)\int_{t'=0}^{t} f_{\text{ext}}(t')\{1 - \exp[(t-t')/\tau_0]\}\mathrm{d}t'$$

$$= (1/m)\int_0^t f_{\text{ext}}(t')\mathrm{d}t' - (1/m)\exp(t/\tau_0)\int_0^t f_{\text{ext}}(t')\exp(-t'/\tau_0)\,\mathrm{d}t'. \qquad (38)$$

The second term on the right-hand side of Eq.(38) is the runaway solution. Dirac [2] suggested to combat this runaway behavior by adding the homogeneous solution $v_1\exp(t/\tau_0)$ to the solution $v(t)$ appearing in Eq.(38). Setting $v_1 = (1/m)\int_0^\infty f_{\text{ext}}(t')\exp(-t'/\tau_0)\,\mathrm{d}t'$ now yields

$$v(t) = (1/m)\int_0^t f_{\text{ext}}(t')\mathrm{d}t' + (1/m)\exp(t/\tau_0)\int_t^\infty f_{\text{ext}}(t')\exp(-t'/\tau_0)\,\mathrm{d}t'. \qquad (39)$$

Note that the above solution is valid for $t \ge 0$, whereas for $t \le 0$ we now have $v(t) = v_1\exp(t/\tau_0)$. The general solution is, of course, continuous at $t = 0$. (If need be, we can add a constant velocity $v_0$ to this solution for all times $t$.) A good way to appreciate the nature of the solution given by Eq.(39) is to differentiate $v(t)$ with respect to $t$ to find the particle's acceleration at time $t$, namely,

$$\dot{v}(t) = (1/m\tau_0)\exp(t/\tau_0)\int_t^\infty f_{\text{ext}}(t')\exp(-t'/\tau_0)\,\mathrm{d}t' = (1/m\tau_0)\int_{\tau=0}^\infty f_{\text{ext}}(t+\tau)\exp(-\tau/\tau_0)\,\mathrm{d}\tau. \qquad (40)$$

Considering that $\tau_0$ is an extremely short time, it is seen from Eq.(40) that the product $m\dot{v}(t)$ of the particle's mass and acceleration is *not* equal to $f_{\text{ext}}(t)$, as one would have guessed from Newton's second law; rather, it is equal to a weighted average of $f_{\text{ext}}(t)$ over a short time interval immediately following $t$. This characteristic of Dirac's solution is referred to as pre-acceleration, since the particle will have to anticipate future values of the applied force in order to adjust its current trajectory. As mentioned earlier, another acausal feature of Dirac's solution is the need for the particle to acquire a velocity $v(t) = v_1\exp(t/\tau_0)$ during the period $t \le 0$, i.e., before the external force $f_{\text{ext}}(t)$ appears on the scene.

**7. Further characteristics of the self-force**. Let us examine the sum of $\dot{\boldsymbol{p}}_{\text{EM}}(t)$, given by Eq.(8), and $\boldsymbol{f}_{\text{self}}(t)$, given by Eq.(27). We have

$$\dot{\boldsymbol{p}}_{\text{EM}}(t) + \boldsymbol{f}_{\text{self}}(t) = \frac{2}{3}\left(\frac{q^2}{4\pi\varepsilon_0 c^2}\right)\left\{\frac{[\dot{\boldsymbol{\beta}}(t)\cdot\dot{\boldsymbol{\beta}}(t)]\boldsymbol{\beta}(t)}{(1-\beta^2)^2} + \frac{[\boldsymbol{\beta}(t)\cdot\dot{\boldsymbol{\beta}}(t)]^2\boldsymbol{\beta}(t)}{(1-\beta^2)^3}\right\}$$

$$+\frac{2}{3}\left(\frac{q^2}{4\pi\varepsilon_0 c^2}\right)\left\{\frac{\ddot{\boldsymbol{\beta}}(t)}{1-\beta^2} + \frac{[\boldsymbol{\beta}(t)\cdot\ddot{\boldsymbol{\beta}}(t)]\boldsymbol{\beta}(t)}{(1-\beta^2)^2} + \frac{3[\boldsymbol{\beta}(t)\cdot\dot{\boldsymbol{\beta}}(t)]\dot{\boldsymbol{\beta}}(t)}{(1-\beta^2)^2} + \frac{3[\boldsymbol{\beta}(t)\cdot\dot{\boldsymbol{\beta}}(t)]^2\boldsymbol{\beta}(t)}{(1-\beta^2)^3}\right\}. \qquad (41)$$

Our goal is to demonstrate that the expression appearing on the right-hand side of Eq.(41) is a complete differential. To this end, we write

$$\frac{\mathrm{d}}{\mathrm{d}t}\left(\frac{\dot{\boldsymbol{\beta}}}{1-\beta^2}\right) = \frac{\ddot{\boldsymbol{\beta}}(t)}{1-\beta^2} + \frac{2(\boldsymbol{\beta}\cdot\dot{\boldsymbol{\beta}})\dot{\boldsymbol{\beta}}}{(1-\beta^2)^2}. \qquad (42)$$

$$\frac{\mathrm{d}}{\mathrm{d}t}\left[\frac{(\boldsymbol{\beta}\cdot\dot{\boldsymbol{\beta}})\boldsymbol{\beta}}{(1-\beta^2)^2}\right] = \frac{(\dot{\boldsymbol{\beta}}\cdot\dot{\boldsymbol{\beta}} + \boldsymbol{\beta}\cdot\ddot{\boldsymbol{\beta}})\boldsymbol{\beta} + (\boldsymbol{\beta}\cdot\dot{\boldsymbol{\beta}})\dot{\boldsymbol{\beta}}}{(1-\beta^2)^2} + \frac{4(\boldsymbol{\beta}\cdot\dot{\boldsymbol{\beta}})^2\boldsymbol{\beta}}{(1-\beta^2)^3}. \qquad (43)$$

Comparing Eq.(41) with the sum of Eqs.(42) and (43), it is easy to see that

$$\dot{\boldsymbol{p}}_{\text{EM}}(t) + \boldsymbol{f}_{\text{self}}(t) = \frac{2}{3}\left(\frac{q^2}{4\pi\varepsilon_0 c^2}\right)\frac{\mathrm{d}}{\mathrm{d}t}\left\{\frac{\dot{\boldsymbol{\beta}}(t)}{1-\beta^2} + \frac{[\boldsymbol{\beta}(t)\cdot\dot{\boldsymbol{\beta}}(t)]\boldsymbol{\beta}(t)}{(1-\beta^2)^2}\right\}. \qquad (44)$$

Integrating Eq.(44) from $t_1$ to $t_2$ reveals that $\int_{t_1}^{t_2}\dot{\boldsymbol{p}}_{\text{EM}}(t)\mathrm{d}t = -\int_{t_1}^{t_2}\boldsymbol{f}_{\text{self}}(t)\,\mathrm{d}t$ under these circumstances:

i) $\dot{\boldsymbol{\beta}}(t_1) = \dot{\boldsymbol{\beta}}(t_2) = 0$; for instance, when the velocities before $t_1$ and after $t_2$ are at constant plateaus, namely, $\boldsymbol{\beta}(t \le t_1) = \boldsymbol{v}_1/c$ and $\boldsymbol{\beta}(t \ge t_2) = \boldsymbol{v}_2/c$.

ii) $\boldsymbol{\beta}(t_1) = \boldsymbol{\beta}(t_2)$ and also $\dot{\boldsymbol{\beta}}(t_1) = \dot{\boldsymbol{\beta}}(t_2)$; for example, when the motion is periodic with period $T = t_2 - t_1$.

A similar analysis can be carried out for the rate of energy radiation, where we now consider the sum of $\dot{\mathcal{E}}_{\mathrm{EM}}(t)$, given by Eq.(7), and $\boldsymbol{f}_{\mathrm{self}}(t)\cdot\boldsymbol{v}(t)$, with $\boldsymbol{f}_{\mathrm{self}}(t)$ given by Eq.(22). We find

$$\dot{\mathcal{E}}_{\mathrm{EM}}(t)+\boldsymbol{f}_{\mathrm{self}}(t)\cdot\boldsymbol{v}(t)=\tfrac{2}{3}\Big(\frac{q^2}{4\pi\varepsilon_0 c}\Big)\Big\{\frac{\dot{\boldsymbol{\beta}}(t)\cdot\dot{\boldsymbol{\beta}}(t)}{(1-\beta^2)^2}+\frac{[\boldsymbol{\beta}(t)\cdot\dot{\boldsymbol{\beta}}(t)]^2}{(1-\beta^2)^3}\Big\}$$

$$+\tfrac{2}{3}\Big(\frac{q^2}{4\pi\varepsilon_0 c}\Big)\Big\{\frac{\ddot{\boldsymbol{\beta}}(t)+[\dot{\boldsymbol{\beta}}(t)\times\boldsymbol{\beta}(t)]\times\dot{\boldsymbol{\beta}}(t)}{(1-\beta^2)^2}+\frac{3[\boldsymbol{\beta}(t)\cdot\dot{\boldsymbol{\beta}}(t)]\{\dot{\boldsymbol{\beta}}(t)+[\boldsymbol{\beta}(t)\times\dot{\boldsymbol{\beta}}(t)]\times\boldsymbol{\beta}(t)\}}{(1-\beta^2)^3}\Big\}\cdot\boldsymbol{\beta}(t)$$

$$=\tfrac{2}{3}\Big(\frac{q^2}{4\pi\varepsilon_0 c}\Big)\Big\{\frac{\dot{\boldsymbol{\beta}}(t)\cdot\dot{\boldsymbol{\beta}}(t)+\ddot{\boldsymbol{\beta}}(t)\cdot\boldsymbol{\beta}(t)}{(1-\beta^2)^2}+\frac{4[\boldsymbol{\beta}(t)\cdot\dot{\boldsymbol{\beta}}(t)]^2}{(1-\beta^2)^3}\Big\}=\tfrac{2}{3}\Big(\frac{q^2}{4\pi\varepsilon_0 c}\Big)\frac{\mathrm{d}}{\mathrm{d}t}\Big[\frac{\boldsymbol{\beta}(t)\cdot\dot{\boldsymbol{\beta}}(t)}{(1-\beta^2)^2}\Big]. \qquad (45)$$

As before, the integration of Eq.(45) reveals that $\int_{t_1}^{t_2}\dot{\mathcal{E}}_{\mathrm{EM}}(t)\mathrm{d}t=-\int_{t_1}^{t_2}\boldsymbol{f}_{\mathrm{self}}(t)\cdot\boldsymbol{v}(t)\,\mathrm{d}t$ under the same conditions (i) and (ii) as mentioned earlier.

**8. Particle in rotary motion**. Consider a point-particle rotating at the constant angular velocity $\boldsymbol{\Omega}=\Omega\hat{\boldsymbol{z}}$ around a circle in the $xy$-plane. The particle velocity $\boldsymbol{v}(t)=c\boldsymbol{\beta}(t)$ along the azimuthal $\hat{\boldsymbol{\varphi}}$-axis has a constant magnitude but changes direction at a constant rate, while its acceleration and its derivative of acceleration (i.e., jerk) are given by $\dot{\boldsymbol{v}}(t)=c\Omega\hat{\boldsymbol{z}}\times\boldsymbol{\beta}(t)$ and $\ddot{\boldsymbol{v}}(t)=-c\Omega^2\boldsymbol{\beta}(t)$, respectively. The rate of energy radiation is thus given by Eq.(7), as follows:

$$\dot{\mathcal{E}}_{\mathrm{EM}}(t)=\tfrac{2}{3}\Big(\frac{q^2}{4\pi\varepsilon_0 c}\Big)\frac{\dot{\boldsymbol{\beta}}(t)\cdot\dot{\boldsymbol{\beta}}(t)}{(1-\beta^2)^2}=\tfrac{2}{3}\Big(\frac{q^2}{4\pi\varepsilon_0 c}\Big)\frac{\Omega^2\beta^2(t)}{(1-\beta^2)^2}. \qquad (46)$$

Considering that, in the present situation, the stored energy in the EM field around the rotating particle cannot vary with time, it should come as no surprise that $\boldsymbol{f}_{\mathrm{self}}(t)\cdot\boldsymbol{v}(t)$ turns out to be equal in magnitude and opposite in sign to the above rate of radiation of energy, that is,

$$\boldsymbol{f}_{\mathrm{self}}(t)\cdot\boldsymbol{v}(t)=\tfrac{2}{3}\Big(\frac{q^2}{4\pi\varepsilon_0 c}\Big)\frac{\ddot{\boldsymbol{\beta}}(t)\cdot\boldsymbol{\beta}(t)}{(1-\beta^2)^2}=-\tfrac{2}{3}\Big(\frac{q^2}{4\pi\varepsilon_0 c}\Big)\frac{\Omega^2\beta^2(t)}{(1-\beta^2)^2}. \qquad (47)$$

As for the time-rate-of-ejection of the EM momentum, we have, from Eq.(8),

$$\dot{\boldsymbol{p}}_{\mathrm{EM}}(t)=\tfrac{2}{3}\Big(\frac{q^2}{4\pi\varepsilon_0 c^2}\Big)\frac{[\dot{\boldsymbol{\beta}}(t)\cdot\dot{\boldsymbol{\beta}}(t)]\boldsymbol{\beta}(t)}{(1-\beta^2)^2}=\tfrac{2}{3}\Big(\frac{q^2}{4\pi\varepsilon_0 c^2}\Big)\frac{\Omega^2\beta^2(t)\boldsymbol{\beta}(t)}{(1-\beta^2)^2}. \qquad (48)$$

However, the self-force is *not* exactly equal (and opposite) to the rate of EM momentum ejection given by Eq.(8), because, in accordance with Eq.(27),

$$\boldsymbol{f}_{\mathrm{self}}(t)=\tfrac{2}{3}\Big(\frac{q^2}{4\pi\varepsilon_0 c^2}\Big)\frac{\ddot{\boldsymbol{\beta}}(t)}{(1-\beta^2)^2}=-\tfrac{2}{3}\Big(\frac{q^2}{4\pi\varepsilon_0 c^2}\Big)\frac{\Omega^2\boldsymbol{\beta}(t)}{(1-\beta^2)^2}. \qquad (49)$$

The preceding equations indicate that the ever-present (intrinsic) momentum in the EM field surrounding the rotating particle must be changing at a rate that equals the difference between Eqs.(48) and (49), namely,

$$\dot{\boldsymbol{p}}_{\mathrm{EM}}^{(\mathrm{intrinsic})}(t)=\tfrac{2}{3}\Big(\frac{q^2}{4\pi\varepsilon_0 c^2}\Big)\frac{\Omega^2\boldsymbol{\beta}(t)}{1-\beta^2}. \qquad (50)$$

The intrinsic EM momentum of a charged particle moving at a constant linear velocity is, of course, aligned with its direction of motion $\boldsymbol{\beta}$. For a particle rotating at a constant angular velocity $\Omega\hat{\boldsymbol{z}}$, one would expect the intrinsic EM momentum to be constant in magnitude and oriented (primarily) along the direction of $\boldsymbol{\beta}(t)$, with the centripetal force accounting for the time-rate-of-change of the particle's linear momentum—which has inertial as well as EM components. Equation (50), however, indicates that the intrinsic EM momentum of the rotating particle has a component in the radial direction as well. According to Eq.(50), the magnitude of this radial component of the particle's EM linear momentum is $p_{\mathrm{EM}}^{(\mathrm{intrinsic})}=\gamma^2q^2\Omega\beta/(6\pi\varepsilon_0c^2)$, which points radially outward from the center of rotation.

**9. The effective Landau-Lifshitz self-force acting on an accelerated point-charge**. At non-relativistic velocities, Newton's law of motion for a point-particle of charge $q$ and mass $m$, following a trajectory $\boldsymbol{r}_p(t)$ under the influence of an external force $\boldsymbol{f}_{\text{ext}}(t)$ and radiation resistance force $\boldsymbol{f}_{\text{self}}(t) \cong q^2\dddot{\boldsymbol{r}}_p(t)/(6\pi\varepsilon_0 c^3)$ is written as follows:

$$m\ddot{\boldsymbol{r}}_p(t) \cong \boldsymbol{f}_{\text{ext}}(t) + q^2\dddot{\boldsymbol{r}}_p(t)/(6\pi\varepsilon_0 c^3). \tag{51}$$

For an electron, the coefficient $\tau_0 = q^2/(6\pi\varepsilon_0 mc^3) \cong 6.24 \times 10^{-24}$ sec is sufficiently small that one may, to a first approximation, ignore the radiation reaction force and write $\dddot{\boldsymbol{r}}_p(t) \cong \dot{\boldsymbol{f}}_{\text{ext}}(t)/m$, in which case Eq.(51) is further approximated as

$$m\ddot{\boldsymbol{r}}_p(t) \cong \boldsymbol{f}_{\text{ext}}(t) + \tau_0\dot{\boldsymbol{f}}_{\text{ext}}(t). \tag{52}$$

If the externally-applied force happens to be a function of both $\boldsymbol{r}$ and $t$, the total time derivative of $\boldsymbol{f}_{\text{ext}}(\boldsymbol{r},t)$ becomes the convective derivative $\partial_t\boldsymbol{f}_{\text{ext}} + (\boldsymbol{v}\cdot\boldsymbol{\nabla})\boldsymbol{f}_{\text{ext}}$, where $\boldsymbol{v} = \dot{\boldsymbol{r}}_p(t)$ is the instantaneous velocity of the point-particle.

When the external force is produced by the electromagnetic fields $\boldsymbol{E}(\boldsymbol{r},t)$ and $\boldsymbol{B}(\boldsymbol{r},t)$ in accordance with the Lorentz force law, namely, $\boldsymbol{f}_{\text{ext}}(t) = q[\boldsymbol{E}(\boldsymbol{r},t) + \dot{\boldsymbol{r}}_p(t)\times\boldsymbol{B}(\boldsymbol{r},t)]$, then, in the particle's instantaneous rest-frame, where $\dot{\boldsymbol{r}}_p(t) = 0$, we will have $\dot{\boldsymbol{f}}_{\text{ext}} = q\dot{\boldsymbol{E}} + q\ddot{\boldsymbol{r}}_p(t)\times\boldsymbol{B}$. Ignoring, as before, the effects of radiation damping in Eq.(51), we find $\ddot{\boldsymbol{r}}_p(t) \cong \boldsymbol{f}_{\text{ext}}(t)/m = q\boldsymbol{E}/m$, which results in the following expression for the radiation reaction force:

$$\boldsymbol{f}_{\text{self}}(t) \cong \tau_0\dot{\boldsymbol{f}}_{\text{ext}}(t) \cong \tau_0 q\dot{\boldsymbol{E}}(t) + (\tau_0 q^2/m)\boldsymbol{E}(t)\times\boldsymbol{B}(t). \quad \leftarrow \boxed{\text{in the rest-frame}} \tag{53}$$

To extend the preceding results to relativistic velocities, we begin by noting that the *proper-time*-derivative of a 4-force acting on a point-particle, although a 4-vector, is *not* a 4-force in and of itself, the reason being that the 4-force $\mathbb{f}(t) = [\gamma\boldsymbol{f}(t)\cdot\boldsymbol{v}(t)/c, \gamma\boldsymbol{f}(t)]$ acting on a point-particle must be orthogonal to the particle's 4-velocity $\mathbb{v}(t) = [\gamma c, \gamma\boldsymbol{v}(t)]$, whereas the derivative (with respect to the proper time $s$) of $\mathbb{f}(t)$ fails to satisfy this constraint. To see this, note that

$$\mathrm{d}\mathbb{f}(t)/\mathrm{d}s = \gamma\big[\gamma^3(\boldsymbol{\beta}\cdot\dot{\boldsymbol{\beta}})\boldsymbol{f}(t)\cdot\boldsymbol{\beta} + \gamma\dot{\boldsymbol{f}}(t)\cdot\boldsymbol{\beta} + \gamma\boldsymbol{f}(t)\cdot\dot{\boldsymbol{\beta}},\ \gamma^3(\boldsymbol{\beta}\cdot\dot{\boldsymbol{\beta}})\boldsymbol{f}(t) + \gamma\dot{\boldsymbol{f}}(t)\big]. \tag{54}$$

$\boxed{\gamma(t) = 1/\sqrt{1-\beta^2(t)}}$ $\boxed{\boldsymbol{\beta}(t) = \boldsymbol{v}(t)/c}$

Clearly, the inner-product of $\mathrm{d}\mathbb{f}(t)/\mathrm{d}s$ and $\mathbb{v}(t)$, namely, $c\gamma^3\boldsymbol{f}(t)\cdot\dot{\boldsymbol{\beta}}(t)$ does not automatically vanish. However, if we form a new 4-vector $\mathbb{g}(t) = \mathrm{d}\mathbb{f}(t)/\mathrm{d}s - (\gamma^3/c)[\boldsymbol{f}(t)\cdot\dot{\boldsymbol{\beta}}(t)]\mathbb{v}(t)$, given that the inner-product of $\mathbb{v}(t)$ with itself is $c^2$, we find that $\mathbb{g}(t)$ is orthogonal to $\mathbb{v}(t)$, as a true 4-force must be. Moreover, in the instantaneous rest-frame of the particle, the space-component of the term that has been subtracted from $\mathrm{d}\mathbb{f}(t)/\mathrm{d}s$ goes to zero, leaving the 3-force component of $\mathbb{g}(t)$ simply as $\dot{\boldsymbol{f}}(s)$. The covariant Landau-Lifshitz radiation reaction 4-force is, therefore, given by $\tau_0\mathbb{g}(t)$, of which the 3-force component is readily found to be (see Eq.(54))

$$\boldsymbol{f}_{\text{self}}(t) \cong \tau_0\gamma\dot{\boldsymbol{f}}_{\text{ext}}(t) + \tau_0\gamma^3(\boldsymbol{\beta}\cdot\dot{\boldsymbol{\beta}})\boldsymbol{f}_{\text{ext}}(t) - \tau_0\gamma^3[\boldsymbol{f}_{\text{ext}}(t)\cdot\dot{\boldsymbol{\beta}}(t)]\boldsymbol{\beta}(t). \tag{55}$$

Now, from the relativistic version of Eq.(51), in which $mc\,\mathrm{d}(\gamma\boldsymbol{\beta})/\mathrm{d}t$ replaces $m\ddot{\boldsymbol{r}}_p(t)$, upon deleting the negligibly small radiation reaction term, we find

$$\gamma^3(\boldsymbol{\beta}\cdot\dot{\boldsymbol{\beta}})\boldsymbol{\beta}(t) + \gamma\dot{\boldsymbol{\beta}}(t) \cong \boldsymbol{f}_{\text{ext}}(t)/mc. \tag{56}$$

The above equation yields

$$\gamma^3(\boldsymbol{\beta}\cdot\dot{\boldsymbol{\beta}})\beta^2 + \gamma(\boldsymbol{\beta}\cdot\dot{\boldsymbol{\beta}}) \cong \boldsymbol{\beta}(t)\cdot\boldsymbol{f}_{\text{ext}}(t)/mc \ \rightarrow\ \gamma^3(\boldsymbol{\beta}\cdot\dot{\boldsymbol{\beta}}) \cong \boldsymbol{\beta}(t)\cdot\boldsymbol{f}_{\text{ext}}(t)/mc. \tag{57}$$

$$\gamma^3(\boldsymbol{\beta}\cdot\dot{\boldsymbol{\beta}})\boldsymbol{\beta}(t)\cdot\boldsymbol{f}_{\text{ext}}(t) + \gamma\dot{\boldsymbol{\beta}}(t)\cdot\boldsymbol{f}_{\text{ext}}(t) \cong \boldsymbol{f}_{\text{ext}}(t)\cdot\boldsymbol{f}_{\text{ext}}(t)/mc$$

$$\rightarrow\ \gamma\dot{\boldsymbol{\beta}}(t)\cdot\boldsymbol{f}_{\text{ext}}(t) \cong [\boldsymbol{f}_{\text{ext}}\cdot\boldsymbol{f}_{\text{ext}} - (\boldsymbol{\beta}\cdot\boldsymbol{f}_{\text{ext}})^2]/mc. \tag{58}$$

Upon substitution of the above expressions into Eq.(55), we find

$$\boldsymbol{f}_{\text{self}}(t) \cong \tau_0\gamma\dot{\boldsymbol{f}}_{\text{ext}}(t) + (\tau_0/mc)(\boldsymbol{\beta}\cdot\boldsymbol{f}_{\text{ext}})\boldsymbol{f}_{\text{ext}}(t) - (\tau_0/mc)\gamma^2(\boldsymbol{f}_{\text{ext}}\cdot\boldsymbol{f}_{\text{ext}})\boldsymbol{\beta}(t)$$
$$+(\tau_0/mc)\gamma^2(\boldsymbol{\beta}\cdot\boldsymbol{f}_{\text{ext}})^2\boldsymbol{\beta}(t). \tag{59}$$

For an externally-applied electromagnetic force $\boldsymbol{f}_{\text{ext}}(t) = q[\boldsymbol{E}(\boldsymbol{r},t) + c\boldsymbol{\beta}(t)\times\boldsymbol{B}(\boldsymbol{r},t)]$, we will have

$$\boldsymbol{\beta}\cdot\boldsymbol{f}_{\text{ext}}(t) = q\boldsymbol{\beta}(t)\cdot\boldsymbol{E}(\boldsymbol{r},t). \tag{60}$$

$$\dot{\boldsymbol{f}}_{\text{ext}}(t) = q[\partial_t\boldsymbol{E} + c(\boldsymbol{\beta}\cdot\boldsymbol{\nabla})\boldsymbol{E}] + qc\dot{\boldsymbol{\beta}}\times\boldsymbol{B} + qc\boldsymbol{\beta}\times[\partial_t\boldsymbol{B} + c(\boldsymbol{\beta}\cdot\boldsymbol{\nabla})\boldsymbol{B}]. \tag{61}$$

$$\gamma\dot{\boldsymbol{\beta}}\times\boldsymbol{B} = \boldsymbol{f}_{\text{ext}}(t)\times\boldsymbol{B}/(mc) - \gamma^3(\boldsymbol{\beta}\cdot\dot{\boldsymbol{\beta}})\boldsymbol{\beta}\times\boldsymbol{B} = [\boldsymbol{f}_{\text{ext}}(t)\times\boldsymbol{B} - (\boldsymbol{\beta}\cdot\boldsymbol{f}_{\text{ext}})\boldsymbol{\beta}\times\boldsymbol{B}]/mc$$
$$= [q\boldsymbol{E}\times\boldsymbol{B} + qc(\boldsymbol{\beta}\times\boldsymbol{B})\times\boldsymbol{B} - (\boldsymbol{\beta}\cdot\boldsymbol{f}_{\text{ext}})\boldsymbol{\beta}\times\boldsymbol{B}]/mc. \tag{62}$$

Consequently,

$$\boldsymbol{f}_{\text{self}}(t) \cong \tau_0\gamma q\{[\partial_t\boldsymbol{E} + c(\boldsymbol{\beta}\cdot\boldsymbol{\nabla})\boldsymbol{E}] + c\boldsymbol{\beta}\times[\partial_t\boldsymbol{B} + c(\boldsymbol{\beta}\cdot\boldsymbol{\nabla})\boldsymbol{B}]\}$$
$$+(\tau_0 q^2/m)[\boldsymbol{E}\times\boldsymbol{B} + c(\boldsymbol{\beta}\times\boldsymbol{B})\times\boldsymbol{B}] - (\tau_0 q/m)(\boldsymbol{\beta}\cdot\boldsymbol{f}_{\text{ext}})\boldsymbol{\beta}\times\boldsymbol{B}$$
$$+(\tau_0/mc)(\boldsymbol{\beta}\cdot\boldsymbol{f}_{\text{ext}})\boldsymbol{f}_{\text{ext}}(t) - (\tau_0/mc)\gamma^2(\boldsymbol{f}_{\text{ext}}\cdot\boldsymbol{f}_{\text{ext}})\boldsymbol{\beta}(t) + (\tau_0/mc)\gamma^2(\boldsymbol{\beta}\cdot\boldsymbol{f}_{\text{ext}})^2\boldsymbol{\beta}(t). \tag{63}$$

Further simplification yields

$$\boldsymbol{f}_{\text{self}}(t) \cong \tau_0 q\gamma\{[\partial_t\boldsymbol{E} + (c\boldsymbol{\beta}\cdot\boldsymbol{\nabla})\boldsymbol{E}] + c\boldsymbol{\beta}\times[\partial_t\boldsymbol{B} + (c\boldsymbol{\beta}\cdot\boldsymbol{\nabla})\boldsymbol{B}]\}$$
$$+(\tau_0 q^2/mc)[(\boldsymbol{\beta}\cdot\boldsymbol{E})\boldsymbol{E} + (\boldsymbol{E} + c\boldsymbol{\beta}\times\boldsymbol{B})\times c\boldsymbol{B}] - (\tau_0/mc)\gamma^2[(\boldsymbol{f}_{\text{ext}}\cdot\boldsymbol{f}_{\text{ext}}) - (\boldsymbol{\beta}\cdot q\boldsymbol{E})^2]\boldsymbol{\beta}. \tag{64}$$

Equation (64) is the covariant expression of the Landau-Lifshitz self-force acting on a point-particle of charge $q$ and mass $m$ in the presence of an external force $\boldsymbol{f}_{\text{ext}}(\boldsymbol{r},t) = q[\boldsymbol{E}(\boldsymbol{r},t) + \dot{\boldsymbol{r}}_p(t)\times\boldsymbol{B}(\boldsymbol{r},t)]$. Note that Eq.(64) reduces to Eq.(53) in the particle's rest-frame.

**10. Concluding remarks**. In this paper, we have shown two methods of calculating the self-force acting on an accelerated point-charge, both of which yield the exact relativistic formula known as the Abraham-Lorentz-Dirac equation. At relativistic speeds of the particle, the method employed in Part 1 of the present paper [1] failed to arrive at this correct result, because it computed the self-force in the laboratory frame by averaging over a *spherical* surface surrounding the point-particle in the limit when the radius of the sphere approached zero. It turns out that, while at non-relativistic speeds the averaging of self-fields over a shrinking spherical surface would result in the correct self-force, the use of a spherical shape at relativistic speeds fails to yield the correct answer. The right approach to solving this problem then is to compute the self-force in the (instantaneous) rest-frame of the particle, then Lorentz transform the result to the laboratory frame, where the particle has a nonzero velocity. This method, paralleling Dirac's original approach [2], was described in Section 3 and yielded the relativistic expression of the self-force in Eq.(27), with details of the calculation relegated to Appendix A. An alternative method of solving the same problem was discussed in Section 2, with the final result given in Eq.(22). In the remainder of the paper, we analyzed certain interesting properties of the general solution obtained in Sections 2 and 3, and also provided a detailed derivation of an approximate form of the self-force first proposed by Landau and Lifshitz [5].

In Part 1 of the present paper [1], our first method of calculating the scalar and vector potentials for an accelerated point-charge yielded expressions for $\psi(\boldsymbol{r},t)$ and $\boldsymbol{A}(\boldsymbol{r},t)$ that were *not* valid at relativistic speeds. Subsequently, we resorted to a method based on Fourier transformation of the charge- and current-density distributions that yielded the correct potentials (i.e., the Liénard-Wiechert potentials [3,4]). At the time of publication of [1], we could not explain why the two methods of calculating the potentials yielded divergent results, but speculated that the Lorentz-FitzGerald contraction of the particle was responsible for the failure of our first method of calculation at relativistic speeds. Appendix B of the present paper describes an extension of the first method of calculation in [1], which now leads to the correct Liénard-Wiechert potentials.

A recurring question asked by students of classical electrodynamics is: "Why is it that a circular loop carrying a constant electrical current produces only a static magnetic field in its surrounding space, even though individual charges that are in circular motion around the loop undergo constant acceleration and must, therefore, continually radiate electromagnetic energy?" We address this question in Appendix C, where we show the connection between the Liénard-Wiechert potentials associated with accelerated charges, and the simpler potentials associated with static charges and constant currents.

**Acknowledgement**. The author is grateful to Vladimir Hnizdo for illuminating discussions on the subjects of this paper.

## Appendix A
## Differentiation with respect to time in special relativity

**A1. Lorentz transformation of the $xyzt$ coordinates**. Let the $x'y'z't'$ frame move with constant velocity $\boldsymbol{v} = v\hat{\boldsymbol{x}}$ along the $x$-axis of the $xyzt$ frame. The corresponding axes of the two inertial frames are parallel to each other, and their origins coincide at $t = t' = 0$. The Lorentz transformation equations [8,9] relating the $(ct', x', y', z')$ coordinates of an event seen in the primed frame to the $(ct, x, y, z)$ coordinates of the same event as seen in the unprimed frame are

$$x = \gamma(x' + vt'); \qquad y = y'; \qquad z = z'; \qquad t = \gamma(t' + vx'/c^2). \tag{A1}$$

In the above equations, $c$ is the speed of light in vacuum, and $\gamma = [1 - (v/c)^2]^{-½}$ is the Lorentz-FitzGerald contraction (or dilation) factor. Inverting the transformation yields the primed coordinates in terms of the unprimed coordinates, as follows:

$$x' = \gamma(x - vt); \qquad y' = y; \qquad z' = z; \qquad t' = \gamma(t - vx/c^2). \tag{A2}$$

It is common to express the Lorentz transformation equations (A1) and (A2) in terms of the normalized velocity $\boldsymbol{\beta} = \boldsymbol{v}/c$, in which case $\gamma$ is written as $1/\sqrt{1-\beta^2}$, where $\beta^2 = \boldsymbol{\beta} \cdot \boldsymbol{\beta}$.

**A2. Transformation of velocities**. Suppose a point-particle has velocity $\boldsymbol{u}' = u'_x\hat{\boldsymbol{x}}' + u'_y\hat{\boldsymbol{y}}' + u'_z\hat{\boldsymbol{z}}'$ when seen in the $x'y'z't'$ frame. The components of the velocity $\boldsymbol{u} = u_x\hat{\boldsymbol{x}} + u_y\hat{\boldsymbol{y}} + u_z\hat{\boldsymbol{z}}$ seen in the $xyzt$ frame are readily computed with the aid of Eq.(A1), as follows:

$$\frac{\Delta x}{\Delta t} = \frac{\gamma_v(\Delta x' + v\Delta t')}{\gamma_v(\Delta t' + v\Delta x'/c^2)} = \frac{(\Delta x'/\Delta t') + v}{1 + (v/c^2)(\Delta x'/\Delta t')} \quad \rightarrow \quad u_x = \frac{u'_x + v}{1 + (u'_x v/c^2)}. \tag{A3a}$$

$$\frac{\Delta y}{\Delta t} = \frac{\Delta y'}{\gamma_v(\Delta t' + v\Delta x'/c^2)} = \frac{\Delta y'/\Delta t'}{\gamma_v[1 + (v/c^2)(\Delta x'/\Delta t')]} \quad \rightarrow \quad u_y = \frac{u'_y}{\gamma_v[1 + (u'_x v/c^2)]}. \tag{A3b}$$

$$\frac{\Delta z}{\Delta t} = \frac{\Delta z'}{\gamma_v(\Delta t' + v\Delta x'/c^2)} = \frac{\Delta z'/\Delta t'}{\gamma_v[1 + (v/c^2)(\Delta x'/\Delta t')]} \quad \rightarrow \quad u_z = \frac{u'_z}{\gamma_v[1 + (u'_x v/c^2)]}. \tag{A3c}$$

**A3. The energy-momentum 4-vector**. Given a point-particle of mass $m$ and velocity $\boldsymbol{u}$, its relativistic energy $\mathcal{E} = \gamma_u mc^2 = mc^2/\sqrt{1-(u/c)^2}$ and linear momentum $\boldsymbol{p} = \gamma_u m\boldsymbol{u}$ form a 4-vector $(\mathcal{E}/c, p_x, p_y, p_z)$. To derive the Lorentz transformation formulas for this energy-momentum 4-vector, consider the relation between $\gamma_u$ of the particle in the $xyzt$ frame and the corresponding $\gamma_{u'}$ in the $x'y'z't'$ frame. Invoking Eqs.(A3), we write

$$\begin{aligned}
\gamma_u^{-2} &= 1 - (u_x^2 + u_y^2 + u_z^2)^2/c^2 = 1 - \frac{(u'_x + v)^2 + (u'^2_y + u'^2_z)/\gamma_v^2}{c^2[1 + (u'_x v/c^2)]^2} \\
&= \frac{c^2[1 + (u'^2_x v^2/c^4) + 2(u'_x v/c^2)] - [(u'^2_x + v^2 + 2u'_x v) + (u'^2_y + u'^2_z) - (u'^2_y + u'^2_z)(v^2/c^2)]}{c^2[1 + (u'_x v/c^2)]^2} \\
&= \frac{(c^2 - v^2) - (u'^2_x + u'^2_y + u'^2_z)[1 - (v^2/c^2)]}{c^2[1 + (u'_x v/c^2)]^2} = \frac{[1 - (v^2/c^2)][1 - (u'^2_x + u'^2_y + u'^2_z)/c^2]}{[1 + (u'_x v/c^2)]^2} = \frac{1}{\gamma_v^2\gamma_{u'}^2[1 + (u'_x v/c^2)]^2}.
\end{aligned} \tag{A4}$$

Relating the energy-momentum 4-vector components in the two frames via Eqs.(A3) and (A4), we now find

$$\begin{aligned}
(\mathcal{E}/c, p_x, p_y, p_z) &= m(\gamma_u c, \gamma_u u_x, \gamma_u u_y, \gamma_u u_z) \\
&= m[\gamma_v\gamma_{u'}(1 + u'_x v/c^2)c,\ \gamma_v\gamma_{u'}(u_{x'} + v),\ \gamma_{u'}u_{y'},\ \gamma_{u'}u_{z'}] \\
&= m[\gamma_v(\gamma_{u'}c + \gamma_{u'}u'_x v/c),\ \gamma_v(\gamma_{u'}u_{x'} + v\gamma_{u'}),\ \gamma_{u'}u_{y'},\ \gamma_{u'}u_{z'}] \\
&= [\gamma_v(\mathcal{E}'/c + vp'_x/c),\ \gamma_v(p'_x + v\mathcal{E}'/c^2),\ p'_y,\ p'_z].
\end{aligned} \tag{A5}$$

The general rule given by Eq.(A5) for transforming the energy-momentum 4-vector between different inertial frames is seen to coincide with that for transforming the space-time coordinates given by Eq.(A1).

**A4. The force 4-vector**: In the $x'y'z't'$ frame, let the point-particle have velocity $\boldsymbol{u}_1'$ at $t_1'$ and velocity $\boldsymbol{u}_2'$ at $t_2' = t_1' + \Delta t'$. The corresponding energy-momentum 4-vectors at both $t_1'$ and $t_2'$ can be Lorentz transformed into the $xyzt$ frame in accordance with Eq.(A5). To transform the force, one must normalize the change of momentum, namely, $\boldsymbol{p}_2' - \boldsymbol{p}_1'$, by $\Delta t'$ in the primed frame, while normalizing $\boldsymbol{p}_2 - \boldsymbol{p}_1$ by $\Delta t$ in the unprimed frame. The trouble is that the particle changes its position between $t_1'$ and $t_2'$, so that $(\Delta x', \Delta y', \Delta z') = (u_x', u_y', u_z')\Delta t'$ is not zero. Consequently, $\Delta t = \gamma_v(\Delta t' + v\Delta x'/c^2) = \gamma_v(1 + vu_x'/c^2)\Delta t'$, which causes the undesirable factor $(1 + vu_x'/c^2)$ to appear in the denominator of the force expression $\boldsymbol{f} = (\boldsymbol{p}_2 - \boldsymbol{p}_1)/\Delta t$ upon dividing $\Delta\boldsymbol{p} = \boldsymbol{p}_2 - \boldsymbol{p}_1$ by $\Delta t$.

To avoid this undesirable situation, we choose the $x'y'z't'$ frame to be the instantaneous rest-frame of the particle [9], i.e., the frame that, at time $t$, travels at the constant velocity $\boldsymbol{u}(t)$ relative to the $xyzt$ frame. The infinitesimal time increment $\Delta t'$ now coincides with the *proper* time increment $\Delta s$ in the particle's (momentary) rest-frame. Since the particle does not move within its rest-frame during $\Delta s$, we now have $\Delta t = \gamma_u \Delta s$. The particle's energy $\mathcal{E}'(s) = \gamma_{u'} mc^2$ inside the rest-frame does not change initially during the short time $\Delta s$, simply because the particle's velocity $\boldsymbol{u}'(s)$ within the rest-frame is zero at first and, therefore,

over-dot indicates differentiation with respect to the proper time $s$. →

$$\dot{\mathcal{E}}'(s) = \dot{\gamma}_{u'} mc^2 = m\gamma_{u'}^3\, \boldsymbol{u}'(s) \cdot \dot{\boldsymbol{u}}'(s) = 0. \tag{A6}$$

As for the particle's momentum $\boldsymbol{p}'(s) = \gamma_{u'} m\boldsymbol{u}'$ inside its momentary rest-frame, we have

$\gamma_{u'}(s) = 1$, $\dot{\gamma}_{u'}(s) = \gamma_{u'}^3\, \boldsymbol{u}'(s) \cdot \dot{\boldsymbol{u}}'(s) = 0$ →

$$\dot{\boldsymbol{p}}'(s) = \dot{\gamma}_{u'} m\boldsymbol{u}'(s) + \gamma_{u'} m\dot{\boldsymbol{u}}'(s) = m\dot{\boldsymbol{u}}'(s). \tag{A7}$$

Thus, inside the (momentary) rest-frame of the particle, the force 4-vector may be defined as $\mathbb{f}'(s) = [\dot{\mathcal{E}}'(s)/c\,, \dot{\boldsymbol{p}}'(s)] = (0, \dot{\boldsymbol{p}}')$. In the laboratory frame, since the differential changes in the energy and momentum are normalized by $\Delta s = \gamma_u^{-1}\Delta t$, the force 4-vector must be described as $\mathbb{f}(t) = [\gamma_u \dot{\mathcal{E}}(t)/c\,, \gamma_u \dot{\boldsymbol{p}}(t)]$, where the over-dot indicates differentiation with respect to the time $t$. Note that, upon Lorentz transformation to the laboratory frame, the rest-frame component of the force that is parallel to the instantaneous direction of motion $\boldsymbol{u}(t)$ of the particle, namely, $\mathbb{f}_\parallel'(s) = m\dot{\boldsymbol{u}}_\parallel'(s) = \boldsymbol{f}_\parallel'(s)$, must become $\gamma_u \boldsymbol{f}_\parallel'(s)$, which then equals the parallel component $\gamma_u \dot{\boldsymbol{p}}_\parallel(t)$ of the 4-focrce in the laboratory frame. Thus, the parallel component of the 3-force in the laboratory frame is found to be $\boldsymbol{f}_\parallel(t) = \dot{\boldsymbol{p}}_\parallel(t) = \boldsymbol{f}_\parallel'(s)$. As for the perpendicular component $\mathbb{f}_\perp'(s) = m\dot{\boldsymbol{u}}_\perp'(s) = \boldsymbol{f}_\perp'(s)$, since it remains the same after the Lorentz transformation, we will have $\boldsymbol{f}_\perp'(s) = \gamma_u \dot{\boldsymbol{p}}_\perp(t)$. The perpendicular 3-force in the laboratory frame is, therefore, given by $\boldsymbol{f}_\perp(t) = \dot{\boldsymbol{p}}_\perp(t) = \gamma_u^{-1}\boldsymbol{f}_\perp'(s)$.

Finally, a Lorentz transform of the temporal component of the 4-force from the rest-frame back to the lab frame yields $\gamma_u \dot{\mathcal{E}}(t)/c = \gamma_u \dot{p}_\parallel'(s)u(t)/c = \gamma_u \dot{p}_\parallel(t)u(t)/c$. Therefore, $\dot{\mathcal{E}}(t) = \dot{p}_\parallel(t)u(t) = \dot{\boldsymbol{p}}(t) \cdot \boldsymbol{u}(t) = \boldsymbol{f}(t) \cdot \boldsymbol{u}(t)$. This, of course, is the same result as one would get by direct differentiation of $\mathcal{E}(t) = \gamma_u mc^2$ and $\boldsymbol{p}(t) = \gamma_u m\boldsymbol{u}(t)$, which yields

$$\dot{\mathcal{E}}(t) = \gamma_u^3 m\boldsymbol{u}(t) \cdot \dot{\boldsymbol{u}}(t). \tag{A8}$$

$$\boldsymbol{f}(t) \cdot \boldsymbol{u}(t) = \dot{\boldsymbol{p}}(t) \cdot \boldsymbol{u}(t) = [\gamma_u^3 m(\boldsymbol{u} \cdot \dot{\boldsymbol{u}})\boldsymbol{u}(t)/c^2 + \gamma_u m\dot{\boldsymbol{u}}(t)] \cdot \boldsymbol{u}(t)$$

$$= \gamma_u^3 m(\boldsymbol{u} \cdot \dot{\boldsymbol{u}})(u^2/c^2 + \gamma_u^{-2}) = \gamma_u^3 m\boldsymbol{u}(t) \cdot \dot{\boldsymbol{u}}(t). \tag{A9}$$

**A5. Generalization to higher time-derivatives**. The result of adding or subtracting two 4-vectors within any given inertial frame is always another 4-vector. Thus, for instance, when we take the time-derivative of the position 4-vector $[ct, \boldsymbol{r}_p(t)]$ of a point-particle, the infinitesimal change in the position 4-vector retains its 4-vector nature. However, normalization by $\mathrm{d}t$ destroys this property of the infinitesimal differential. Nevertheless, if we normalize the differential of the 4-vector by the differential $\mathrm{d}s$ of the "proper time" (as opposed to the differential $\mathrm{d}t$ of the laboratory time), then the properly normalized time-derivative retains a 4-vector identity [9]. Considering that $\mathrm{d}t = \gamma \mathrm{d}s$, it is readily seen that the product of $\gamma$ and the time-derivative of any 4-vector entity is a legitimate 4-vector in its own right. Thus, whereas the derivative $[c, \dot{\boldsymbol{r}}_p(t)]$ of the position 4-vector is *not* a 4-vector, its product with $\gamma$, namely, $\gamma[c, \dot{\boldsymbol{r}}_p(t)] = [c\gamma, \gamma \boldsymbol{v}(t)] = c[\gamma, \gamma\boldsymbol{\beta}(t)]$, is a 4-vector — commonly referred to as the velocity 4-vector, or the 4-velocity. We may now repeat the process and construct an acceleration 4-vector by multiplying $\gamma$ into the derivative of the velocity 4-vector. Dropping the constant coefficient $c$ (for convenience's sake), recalling that $\gamma = (1-\beta^2)^{-½}$, and denoting the time-derivative (with respect to $t$) with an over-dot, we will have

$$\gamma \frac{\mathrm{d}}{\mathrm{d}t}(\gamma, \gamma\boldsymbol{\beta}) = \gamma(\dot{\gamma}, \dot{\gamma}\boldsymbol{\beta} + \gamma\dot{\boldsymbol{\beta}}) = [\gamma^4 \boldsymbol{\beta}\cdot\dot{\boldsymbol{\beta}},\ \gamma^4(\boldsymbol{\beta}\cdot\dot{\boldsymbol{\beta}})\boldsymbol{\beta} + \gamma^2\dot{\boldsymbol{\beta}}]. \tag{A10}$$

Next, we differentiate (again with respect to $t$) the acceleration 4-vector one more time, and proceed to multiply the result by $\gamma$, to arrive at a new 4-vector for the jerk of the particle, as follows:

$$\gamma \frac{\mathrm{d}}{\mathrm{d}t}[\gamma^4 \boldsymbol{\beta}\cdot\dot{\boldsymbol{\beta}},\ \gamma^4(\boldsymbol{\beta}\cdot\dot{\boldsymbol{\beta}})\boldsymbol{\beta} + \gamma^2\dot{\boldsymbol{\beta}}] = [\overbrace{4\gamma^7(\boldsymbol{\beta}\cdot\dot{\boldsymbol{\beta}})^2 + \gamma^5(\dot{\boldsymbol{\beta}}\cdot\dot{\boldsymbol{\beta}} + \boldsymbol{\beta}\cdot\ddot{\boldsymbol{\beta}})}^{\text{temporal component}},$$
$$\underbrace{4\gamma^7(\boldsymbol{\beta}\cdot\dot{\boldsymbol{\beta}})^2\boldsymbol{\beta} + \gamma^5(\dot{\boldsymbol{\beta}}\cdot\dot{\boldsymbol{\beta}} + \boldsymbol{\beta}\cdot\ddot{\boldsymbol{\beta}})\boldsymbol{\beta} + 3\gamma^5(\boldsymbol{\beta}\cdot\dot{\boldsymbol{\beta}})\dot{\boldsymbol{\beta}} + \gamma^3\ddot{\boldsymbol{\beta}}}_{\text{spatial component}}]. \tag{A11}$$

In the particle's rest-frame (r-f), $\boldsymbol{\beta}_{\mathrm{r-f}} = 0$, $\gamma_{\mathrm{r-f}} = 1$, and the above expression reduces to $(\dot{\boldsymbol{\beta}}_{\mathrm{r-f}}\cdot\dot{\boldsymbol{\beta}}_{\mathrm{r-f}},\ \ddot{\boldsymbol{\beta}}_{\mathrm{r-f}})$. Here $\dot{\boldsymbol{\beta}}_{\mathrm{r-f}}$ and $\ddot{\boldsymbol{\beta}}_{\mathrm{r-f}}$ are, respectively, the acceleration and the jerk of the particle in its rest-frame. To determine $\ddot{\boldsymbol{\beta}}_{\mathrm{r-f}}$, the jerk 4-vector in Eq.(A11) must be Lorentz transformed from the laboratory frame to the particle's rest-frame. The spatial component of jerk that is perpendicular ($\perp$) to the direction of motion is transformed to the rest-frame without any change at all. To determine this component, we cross-multiply, twice, the spatial component of the jerk 4-vector, given by Eq.(A11), into the unit-vector $\boldsymbol{\beta}/\beta$, which yields

$$\ddot{\boldsymbol{\beta}}^{\perp}_{\mathrm{r-f}} = -\boldsymbol{\beta}\times\{\boldsymbol{\beta}\times[3\gamma^5(\boldsymbol{\beta}\cdot\dot{\boldsymbol{\beta}})\dot{\boldsymbol{\beta}} + \gamma^3\ddot{\boldsymbol{\beta}}]\}/\beta^2 = [3\gamma^5(\boldsymbol{\beta}\cdot\dot{\boldsymbol{\beta}})\dot{\boldsymbol{\beta}} + \gamma^3\ddot{\boldsymbol{\beta}}] - [3\gamma^5(\boldsymbol{\beta}\cdot\dot{\boldsymbol{\beta}})^2 + \gamma^3(\boldsymbol{\beta}\cdot\ddot{\boldsymbol{\beta}})]\,\boldsymbol{\beta}/\beta^2. \tag{A12}$$

To transform the parallel component (‖) of the jerk to the rest-frame, we first isolate the spatial component of the jerk 4-vector that is aligned with $\boldsymbol{\beta}/\beta$, as follows:

$$4\gamma^7(\boldsymbol{\beta}\cdot\dot{\boldsymbol{\beta}})^2\boldsymbol{\beta} + \gamma^5(\dot{\boldsymbol{\beta}}\cdot\dot{\boldsymbol{\beta}} + \boldsymbol{\beta}\cdot\ddot{\boldsymbol{\beta}})\boldsymbol{\beta} + 3\gamma^5(\boldsymbol{\beta}\cdot\dot{\boldsymbol{\beta}})^2\,\boldsymbol{\beta}/\beta^2 + \gamma^3(\boldsymbol{\beta}\cdot\ddot{\boldsymbol{\beta}})\,\boldsymbol{\beta}/\beta^2. \tag{A13}$$

Next, we multiply the temporal component of the jerk 4-vector of Eq.(A11) by $\boldsymbol{\beta}$, subtract the result from its spatial parallel component given by Eq.(A13), and multiply the resulting 3-vector by $\gamma$. This yields the parallel component of $\ddot{\boldsymbol{\beta}}_{\mathrm{r-f}}$, as follows:

$$\ddot{\boldsymbol{\beta}}^{\parallel}_{\mathrm{r-f}} = \gamma\{[4\gamma^7(\boldsymbol{\beta}\cdot\dot{\boldsymbol{\beta}})^2\boldsymbol{\beta} + \gamma^5(\dot{\boldsymbol{\beta}}\cdot\dot{\boldsymbol{\beta}} + \boldsymbol{\beta}\cdot\ddot{\boldsymbol{\beta}})\boldsymbol{\beta} + 3\gamma^5(\boldsymbol{\beta}\cdot\dot{\boldsymbol{\beta}})^2\,\boldsymbol{\beta}/\beta^2 + \gamma^3(\boldsymbol{\beta}\cdot\ddot{\boldsymbol{\beta}})\,\boldsymbol{\beta}/\beta^2]$$
$$-[4\gamma^7(\boldsymbol{\beta}\cdot\dot{\boldsymbol{\beta}})^2 + \gamma^5(\dot{\boldsymbol{\beta}}\cdot\dot{\boldsymbol{\beta}} + \boldsymbol{\beta}\cdot\ddot{\boldsymbol{\beta}})]\boldsymbol{\beta}\} = [3\gamma^6(\boldsymbol{\beta}\cdot\dot{\boldsymbol{\beta}})^2\boldsymbol{\beta} + \gamma^4(\boldsymbol{\beta}\cdot\ddot{\boldsymbol{\beta}})\boldsymbol{\beta}]/\beta^2. \tag{A14}$$

According to Eq.(10), the self-force of the EM field acting on the charged particle in its rest-frame is $q^2\ddot{\boldsymbol{\beta}}_{\mathrm{r-f}}/(6\pi\varepsilon_0 c^2)$. The 4-force must be Lorentz transformed back into the laboratory frame. In the process, the perpendicular component of the 4-force remains the same, while its parallel component gets multiplied by $\gamma$. Once in the laboratory frame, one can recover the 3-force by dividing both components (i.e., ‖ as well as $\perp$) of the 4-force by $\gamma$. Subsequently, the self-force of the point-particle in the laboratory frame is found to be

$$\left(\frac{q^2}{6\pi\varepsilon_0 c^2}\right)[\ddot{\boldsymbol{\beta}}^{\parallel}_{\mathrm{r-f}} + \gamma^{-1}\ddot{\boldsymbol{\beta}}^{\perp}_{\mathrm{r-f}}] = \left(\frac{q^2}{6\pi\varepsilon_0 c^2}\right)[3\gamma^6(\boldsymbol{\beta}\cdot\dot{\boldsymbol{\beta}})^2\boldsymbol{\beta} + \gamma^4(\boldsymbol{\beta}\cdot\ddot{\boldsymbol{\beta}})\boldsymbol{\beta} + 3\gamma^4(\boldsymbol{\beta}\cdot\dot{\boldsymbol{\beta}})\dot{\boldsymbol{\beta}} + \gamma^2\ddot{\boldsymbol{\beta}}]. \tag{A15}$$

Equation (A15) is the exact formula for the self-force acting on an accelerated point-charge.

## Appendix B
## The Liénard-Wiechert potentials

In the Lorenz gauge, the scalar potential produced by the electric charge distribution $\rho(\boldsymbol{r},t)$ is given by

$$\psi(\boldsymbol{r},t)=\frac{1}{4\pi\varepsilon_0}\int_{-\infty}^{\infty}\frac{\rho(\boldsymbol{r}',t-|\boldsymbol{r}-\boldsymbol{r}'|/c)}{|\boldsymbol{r}-\boldsymbol{r}'|}\mathrm{d}x'\mathrm{d}y'\mathrm{d}z'. \tag{B1}$$

In Part 1 of the present paper [1], we initially found an incorrect expression for the scalar potential, when we substituted the charge-density $q\delta[x'-x_p(t')]\delta[y'-y_p(t')]\delta[z'-z_p(t')]$ of a traveling point-charge $q$ for $\rho(\boldsymbol{r}',t')$ in the integrand of Eq.(B1), the retarded time being $t'=t-|\boldsymbol{r}-\boldsymbol{r}'|/c$. A similar problem afflicted our initial expression for the vector potential $\boldsymbol{A}(\boldsymbol{r},t)$. Subsequently, we resorted to an alternative method of calculating $\psi(\boldsymbol{r},t)$ and $\boldsymbol{A}(\boldsymbol{r},t)$ for an accelerated point-charge, and arrived at the correct expressions for the Liénard-Wiechert potentials (see Section 3, Part 1). Throughout the rest of the paper [1], we relied on the correct Liénard-Wiechert potentials, but could not pinpoint the root cause of the difference between the two methods of calculating the scalar and vector potentials of an accelerated point-charge. The goal of the present Appendix is to point out the source of the discrepancy between the two methods of calculating the potentials.

If we introduce a 4$^{\text{th}}$ $\delta$-function (i.e., one associated with $t'$) into the integrand of Eq.(B1), we will have

$$\psi(\boldsymbol{r},t)=\frac{q}{4\pi\varepsilon_0}\int_{-\infty}^{\infty}\frac{\delta[x'-x_p(t')]\times\delta[y'-y_p(t')]\times\delta[z'-z_p(t')]\times\delta[t'-t+|\boldsymbol{r}-\boldsymbol{r}'|/c]}{|\boldsymbol{r}-\boldsymbol{r}'|}\mathrm{d}x'\mathrm{d}y'\mathrm{d}z'\mathrm{d}t'. \tag{B2}$$

It is now easy to evaluate the above integral by first integrating over the $x',y',z'$ coordinates, as follows:

$$\psi(\boldsymbol{r},t)=\frac{q}{4\pi\varepsilon_0}\int_{-\infty}^{\infty}\frac{\delta[t'-t+|\boldsymbol{r}-\boldsymbol{r}_p(t')|/c]}{|\boldsymbol{r}-\boldsymbol{r}_p(t')|}\mathrm{d}t'. \tag{B3}$$

To evaluate the remaining integral over $t'$, we invoke the $\delta$-function property that, in the vicinity of each and every zero $t_0'$ of the function $f(t')$, one has $\delta[f(t')]=\delta(t'-t_0')/|\dot{f}(t_0')|$, where $\dot{f}(t_0')$ is the derivative of $f(t')$ at $t'=t_0'$. In the case of the integrand of Eq.(B3), the zero of the argument of the $\delta$-function occurs when $|\boldsymbol{r}-\boldsymbol{r}_p(t')|=c(t-t')$. We now write

$$\partial[t'-t+|\boldsymbol{r}-\boldsymbol{r}_p(t')|/c]/\partial t'=1-[\boldsymbol{r}-\boldsymbol{r}_p(t')]\cdot\dot{\boldsymbol{r}}_p(t')/c|\boldsymbol{r}-\boldsymbol{r}_p(t')|. \tag{B4}$$

Invoking the sifting property of Dirac's $\delta$-function, Eq.(B3) can finally be written as follows:

$$\psi(\boldsymbol{r},t)=\frac{q}{4\pi\varepsilon_0\{|\boldsymbol{r}-\boldsymbol{r}_p(t')|-[\boldsymbol{r}-\boldsymbol{r}_p(t')]\cdot\dot{\boldsymbol{r}}_p(t')/c\}}. \tag{B5}$$

The retarded time $t'$ appearing in the denominator on the right-hand side of Eq.(B5) is the unique solution of the equation $|\boldsymbol{r}-\boldsymbol{r}_p(t')|=c(t-t')$. Equation (B5) is the standard expression of the Liénard-Wiechert scalar potential in the Lorenz gauge. A similar expression may be readily obtained for the vector potential $\boldsymbol{A}(\boldsymbol{r},t)$.

In retrospect, it is now easy to see why the method employed in Section 2 of Part 1 of the present paper failed to reproduce the correct Liénard-Wiechert potentials. When the integral in Eq.(B1) is evaluated over the 3-dimensional $\delta$-function $\delta[\boldsymbol{r}'-\boldsymbol{r}_p(t')]$, it is *not* permissible at relativistic speeds of the particle to resolve this $\delta$-function into the triple product $\delta[x'-x_p(t')]\times\delta[y'-y_p(t')]\times\delta[z'-z_p(t')]$. This is because, if one conceives of the particle as a small spherical ball of charge, then, at relativistic speeds (where $\dot{\boldsymbol{r}}_p(t)$ is comparable to the speed $c$ of light in free space), the Lorentz-FitzGerald contraction causes the deformation of the sphere into a spheroidal shape, at which point the three-dimensional $\delta$-function can no longer be separated into the product of three independent $\delta$-functions along the Cartesian coordinate axes. The method presented in this Appendix shows that the problem disappears when a fourth $\delta$-function (this one associated with the time dimension) is introduced into the integrand, as is done in Eq.(B2).

# Appendix C

## Reconciling the scalar and vector potentials associated with constant electric currents with the Liénard-Wiechert potentials of a moving point charge

**C1. The case of an infinitely long, thin wire carrying a constant current**. With reference to Fig.C1(a), since the electrons are moving at constant velocity $v$ along the $z$-axis, the potentials at the observation point $\boldsymbol{r} = (x_0, 0, 0)$ must be the Liénard-Wiechert potentials given in Eqs.(3) and (4). However, these are *not* the same potentials as those for constant charge- and current-densities along the length of the wire. If, at time $t$, the electrons happen to be inside an infinitesimal segment of length $\mathrm{d}z$ centered at $z$, then they must have contributed to the observed potential at $\boldsymbol{r} = (x_0, 0, 0)$ a short time earlier, when they were located at $z'$. The retardation condition yields

$$
\begin{aligned}
& c(z-z')/v = \sqrt{x_0^2 + z'^2} && \rightarrow \quad z^2 + z'^2 - 2zz' = \beta^2(x_0^2 + z'^2) \\
\rightarrow \quad & (1-\beta^2)z'^2 - 2zz' + (z^2 - \beta^2 x_0^2) = 0 && \rightarrow \quad z'^2 - 2\gamma^2 zz' + \gamma^2(z^2 - \beta^2 x_0^2) = 0 \\
\rightarrow \quad & z' = \gamma^2 z - \sqrt{\gamma^4 z^2 - \gamma^2 z^2 + \gamma^2\beta^2 x_0^2} && \rightarrow \quad z' = \gamma^2\Big[z - \beta\sqrt{z^2 + (x_0/\gamma)^2}\Big].
\end{aligned} \tag{C1}
$$

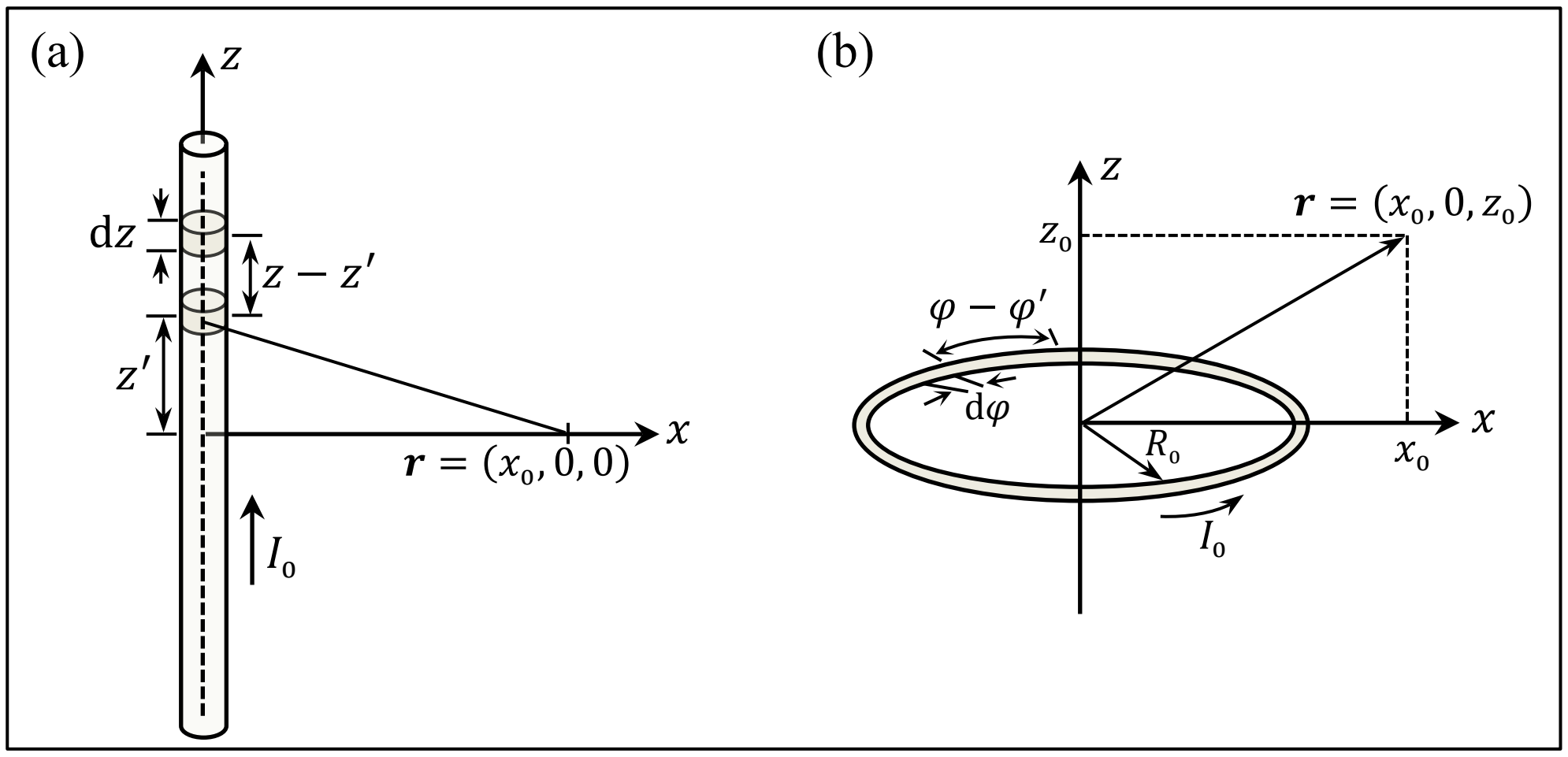

**Fig.C1**. (a) An infinitely-long, thin wire carries the constant current $I_0$ long the $z$-axis. The electrons within a differential element $\mathrm{d}z$ of the wire, located at height $z$ at time $t$, make their contributions to the scalar and vector potentials observed at $\boldsymbol{r} = (x_0, 0, 0)$ at an earlier time when they were at the point $z'$. (b) A thin circular ring of radius $R_0$, placed in the $xy$-plane and centered at the origin of coordinates, carries a constant current $I_0$ in the $\hat{\boldsymbol{\varphi}}$ direction. The electrons inside a differential segment $\mathrm{d}\varphi$ of the wire, located at the azimuthal coordinate $\varphi$ at time $t$, make their contributions to the scalar and vector potentials observed at $\boldsymbol{r} = (x_0, 0, z_0)$ at an earlier time when they were at the point $\varphi'$.

The denominator in the integrand of the Liénard-Wiechert potentials is given by

$$
\begin{aligned}
\sqrt{x_0^2 + z'^2} + \beta z' &= \Big\{x_0^2 + \gamma^4\Big[(1+\beta^2)z^2 + (\beta x_0/\gamma)^2 - 2\beta z\sqrt{z^2 + (x_0/\gamma)^2}\Big]\Big\}^{1/2} + \beta\gamma^2\Big[z - \beta\sqrt{z^2 + (x_0/\gamma)^2}\,\Big] \\
&= \gamma^2\Big[z^2 + (x_0/\gamma)^2 + \beta^2 z^2 - 2\beta z\sqrt{z^2 + (x_0/\gamma)^2}\,\Big]^{1/2} + \gamma^2\Big[\beta z - \beta^2\sqrt{z^2 + (x_0/\gamma)^2}\Big] \\
&= \gamma^2\Big[\sqrt{z^2 + (x_0/\gamma)^2} - \beta z\Big] + \gamma^2\Big[\beta z - \beta^2\sqrt{z^2 + (x_0/\gamma)^2}\,\Big] \\
&= \gamma^2(1-\beta^2)\sqrt{z^2 + (x_0/\gamma)^2} = \gamma^{-1}\sqrt{(\gamma z)^2 + x_0^2}\,.
\end{aligned} \tag{C2}
$$

The integral yielding both the scalar and vector potentials may, therefore, be written as

$$\int_{-\infty}^{\infty}\frac{\mathrm{d}z}{\sqrt{x_0^2+z'^2}+\beta z'}=\int_{-\infty}^{\infty}\frac{\gamma\mathrm{d}z}{\sqrt{(\gamma z)^2+x_0^2}}=\int_{-\infty}^{\infty}\frac{\mathrm{d}\zeta}{\sqrt{\zeta^2+x_0^2}}. \qquad \text{(C3)}$$

The last integral in Eq.(C3) is what one obtains upon direct integration of the charge-density without regard for the motion of the electrons. (The same integral is also obtained for the current-density in the present case.)

An alternative derivation of the above result entails the differentiation of the first line of Eq.(C1) with respect to $z$, which yields

$$\mathrm{d}z-\mathrm{d}z'=\beta\mathrm{d}z'/\sqrt{x_0^2+z'^2} \quad\rightarrow\quad \mathrm{d}z=\left(1+\beta/\sqrt{x_0^2+z'^2}\right)\mathrm{d}z'. \qquad \text{(C4)}$$

Upon substituting the above $\mathrm{d}z$ into the left-hand integral in Eq.(C3), one arrives at the desired result.

**C2. The case of a constant current around a circular ring**. Let a thin, circular ring of radius $R_0$ carry the constant current $I_0$ in the $\hat{\boldsymbol{\varphi}}$ direction, as depicted in Fig.C1(b). Here the moving charges within the infinitesimal element $\mathrm{d}\varphi$ of the ring at time $t$, must have contributed to the potentials observed at $\boldsymbol{r}=(x_0,0,z_0)$ when they were at the retarded position $\varphi'$. We thus have

$$cR_0(\varphi-\varphi')/v=\sqrt{(x_0-R_0\cos\varphi')^2+R_0^2\sin^2\varphi'+z_0^2}=\sqrt{x_0^2+z_0^2+R_0^2-2x_0R_0\cos\varphi'}. \qquad \text{(C5)}$$

Differentiating both sides of Eq.(C5) with respect to $\varphi$, we find

$$R_0(\mathrm{d}\varphi-\mathrm{d}\varphi')/\beta=x_0R_0\sin\varphi'\,\mathrm{d}\varphi'/\sqrt{x_0^2+z_0^2+R_0^2-2x_0R_0\cos\varphi'}. \qquad \text{(C6)}$$

Consequently,

$$\mathrm{d}\varphi=\left(1+\beta x_0\sin\varphi'/\sqrt{x_0^2+z_0^2+R_0^2-2x_0R_0\cos\varphi'}\right)\mathrm{d}\varphi'. \qquad \text{(C7)}$$

It is now easy to see that the denominator of the integrand of the Liénard-Wiechert scalar potential $\psi(\boldsymbol{r},t)$, namely, $|\boldsymbol{r}-\boldsymbol{r}_p(t')|-[\boldsymbol{r}-\boldsymbol{r}_p(t')]\cdot\dot{\boldsymbol{r}}_p(t')/c$, can be streamlined as follows:

$$\begin{aligned}&\sqrt{x_0^2+z_0^2+R_0^2-2x_0R_0\cos\varphi'}-[(x_0-R_0\cos\varphi')\hat{\boldsymbol{x}}-R_0\sin\varphi'\,\hat{\boldsymbol{y}}+z_0\hat{\boldsymbol{z}}]\cdot(-\beta\sin\varphi'\,\hat{\boldsymbol{x}}+\beta\cos\varphi'\,\hat{\boldsymbol{y}})\\&\quad=\sqrt{x_0^2+z_0^2+R_0^2-2x_0R_0\cos\varphi'}+\beta[(x_0-R_0\cos\varphi')\sin\varphi'+R_0\sin\varphi'\cos\varphi']\\&\quad=\sqrt{x_0^2+z_0^2+R_0^2-2x_0R_0\cos\varphi'}+\beta x_0\sin\varphi'. \qquad \text{(C8)}\end{aligned}$$

Consequently,

$$\int_{\varphi=0}^{2\pi}\frac{R_0\mathrm{d}\varphi}{\sqrt{x_0^2+z_0^2+R_0^2-2x_0R_0\cos\varphi'}+\beta x_0\sin\varphi'}=\int_{\varphi'=0}^{2\pi}\frac{R_0\mathrm{d}\varphi'}{\sqrt{x_0^2+z_0^2+R_0^2-2x_0R_0\cos\varphi'}}. \qquad \text{(C9)}$$

The right-hand side of Eq.(C9) is the same expression as obtained for the scalar potential of immobile charges uniformly distributed around a circular ring of radius $R_0$. A similar argument holds for the vector potential $\boldsymbol{A}(\boldsymbol{r},t)$ of the current loop.